\documentclass{JHEP3}
\usepackage{graphics}
\usepackage{epsfig}
\usepackage{axodraw}
\usepackage{feynmp}

\newcommand{\be}{\begin{equation}}
\newcommand{\ee}{\end{equation}}
\newcommand{\bea}{\begin{eqnarray}}
\newcommand{\eea}{\end{eqnarray}}
\newcommand{\bi}{\begin{itemize}}
\newcommand{\ei}{\end{itemize}}
\newcommand{\bc}{\begin{center}}
\newcommand{\ec}{\end{center}}
\newcommand{\bfi}{\begin{figure}}
\newcommand{\efi}{\end{figure}}

\def\psih{{\psi_{\hat\Phi}}}
\def\tev{\,{\rm TeV}}
\def\half{{1\over2}}

\def\Id{{1\!\!\!\!1}}
\def\Tr{{\rm Tr\,}}

\title{Direct Mediation and a Visible Metastable Supersymmetry Breaking Sector}

\author{Boaz Keren-Zur, Luca Mazzucato
  and Yaron Oz\\
Raymond and Beverly Sackler Faculty of Exact Sciences \\
School of Physics and Astronomy \\
Tel-Aviv University, Ramat-Aviv 69978, Israel\\
E-mails: \email{kerenzu@post.tau.ac.il},
  \email{mazzul@post.tau.ac.il},
  \email{yaronoz@post.tau.ac.il}
}

\abstract{We analyze an R-symmetry breaking deformation of the ISS model for a direct mediation of supersymmetry  breaking from a metastable vacuum.
The model is weakly coupled and calculable. The LSP gravitino is light ($m_{3\over2}<16\,{\rm eV}$) and the MSSM  spectrum is natural with a light Higgs.
The supersymmetry breaking sector, which is usually hidden, is observable ($m \sim1 {\,\rm TeV}$) and may be a candidate for cold dark matter. We discuss its production and signature at LHC. We propose a UV completion of the model in terms of a duality cascade.}


\begin{document}

\section{Introduction and summary}

An important question of particle physics
 is the nature of supersymmetry breaking and its
mediation to the MSSM particles.
In the phenomenological approach the main motivation for introducing
supersymmetry is the resolution of the gauge hierarchy problem.
Introduction of the superpartners results in the cancellation of
all quadratic divergences in the theory.
However, this on its own
does not give an explanation for the energy
scale of supersymmetry breaking
and why it is so much smaller than the Planck scale.
The general answer to this question involves an asymptotically free
gauge force (in a hidden sector of the theory, $i.\,e.$ outside the Standard Model)
which becomes strong at low energies,
and then non-perturbative effects trigger spontaneous supersymmetry breakdown.
This mechanism is known as dynamical supersymmetry breaking (DSB).

Until recently, it was presumed that DSB requires that
non-supersymmetric vacuum state of the hidden sector is the true vacuum,
$i.\,e.$ the global minimum of the effective potential. In models where supersymmetry breaking is transmitted to the Standard Model by gauge interactions (a.k.a. gauge mediation), this requirement is hard to satisfy, which made DSB models
and the mediation mechanism to the Standard Model sector rather complicated.
ISS \cite{Intriligator:2006dd}
proposed a simple DSB model in which the non-supersymmetric vacuum state is
metastable with a very low tunneling rate to the true supersymmetric
vacuum.
The ISS model has a large unbroken
flavor symmetry, which can be weakly gauged without spoiling the
DSB mechanism. This makes it a convenient framework for a direct
gauge mediation of supersymmetry breaking to the Standard Model,
where some of the DSB-sector
particles are also charged under the Standard Model
$SU(3)\times SU(2)\times U(1)$ gauge group.
In such models, all the superpartners of Standard Model particles
become massive via 1--loop and 2--loop diagrams, and their masses
are calculable in terms of the DSB sector parameters.

However, in order to build a model of direct gauge mediation, one needs to overcome two features of the ISS model,
which are problematic for phenomenology. The first issue is the presence of an accidental R-symmetry, that forbids the generation of gaugino masses.
The second issue is the spontaneous breaking of the flavor symmetry group, that introduces Goldstone bosons charged under the Standard Model gauge group.
We will resolve both issues by breaking explicitly the R-symmetry and the flavor symmetry by mass terms that deform the ISS model \cite{Kitano:2006xg}.\footnote{Other recent analysis of direct gauge mediation using various deformations of the ISS appear in \cite{Csaki:2006wi}\cite{Haba:2007rj}\cite{Abel:2007jx}\cite{Abel:2007nr}\cite{Dine:2007dz}.}

The model that we will study in this paper is weakly coupled and calculable. The LSP gravitino is light ($m_{3\over2}<16\,{\rm eV}$), as required by the cosmological bounds \cite{Viel:2005qj} for gauge mediation, and the MSSM  spectrum is natural with a light Higgs. In particular, at the expense of genericity, we obtain a model where there is no tension between a long lifetime of the metastable vacuum and a large gaugino/scalar mass ratio, which typically leads to split supersymmetry.
The supersymmetry breaking sector, which is usually hidden, is observable ($m \sim$ 1 TeV). We discuss in detail its features, its production cross section at LHC and some of its decay channels. Moreover, the DSB sector contains possibly long-lived particles that may be a candidate for cold dark matter.

It is generically difficult to avoid a Landau pole in models of direct gauge mediation.
In our model as well, the QCD coupling runs very fast above a certain scale, hitting a Landau pole below the GUT scale. We propose a UV completion in terms of a duality cascade \cite{Klebanov:2000hb}\cite{Heckman:2007zp}\cite{Franco:2008jc} by embedding the MSSM coupled to the supersymmetry breaking sector in a quiver gauge theory.
When the QCD coupling hits the Landau pole, the first step of the duality cascade is triggered and we discuss it.
The perturbative unification in the dual quiver is still an open issue.

The paper is organized as follows.
In section 2 we review the ISS model and its deformation and we identify the metastable vacuum. In section 3 we discuss the various requirements that constrain the parameter space of the model, such as a light gravitino, the absence of tachyons and a long lifetime of the metastable vacuum. In this section
we also discuss the generation of the soft terms in the MSSM. In section 4 we discuss in detail the phenomenology of the light particles coming from the supersymmetry breaking sector, which in our case will be observable at LHC. In section 5, we present the salient features of the MSSM spectrum, after taking into account the RG evolution from the messenger scale down to the TeV scale. In section 6 we propose a particular UV completion of our model, in terms of a duality cascade, by embedding the MSSM coupled to the supersymmetry breaking sector into a quiver gauge theory. There are three appendices in which we outline some calculations.

\section{The supersymmetry breaking vacuum}

 In order to construct a model of direct gauge mediation based on the ISS one, we embed the MSSM gauge groups into the flavor symmetry group of the ISS. However, we need to overcome two of the ISS features which are problematic for phenomenology: the presence of an accidental R-symmetry, that forbids the generation of gaugino masses, and the spontaneous breaking of the flavor symmetry group, that introduces Goldstone bosons charged under the would-be MSSM gauge groups.
We will consider a model that is a deformation of the ISS one, which has been proposed by \cite{Kitano:2006xg}.

\subsection{Deformation of the ISS model}

We will work in the magnetic dual description of ${\cal N}=1$ $SU(N_c)$ SQCD with $N_f$ flavors.
The magnetic gauge group is $SU(N)$ ($N=N_f -N_c$) and we have $N_f$ flavors of (magnetic) quarks and antiquarks $\tilde q^f$ and $q_{\tilde f}$, coupled to $N_f^2$ singlet chiral superfields $\Phi^{\tilde f}_f$, via the superpotential
\be
W=h\Tr \tilde q\Phi q-h\mu^2\Tr\Phi \ ,
\ee
with the second term corresponding to the mass term of the electric quarks.
 This theory has a global $SU(N_f)\times U(1)_B\times U(1)_R$ symmetry, which is spontaneously broken to $SU(N)_{\rm diag}\times SU(N_f-N)\times U(1)_R$ in the ISS vacuum by the expectation value $\tilde{q} q = \mu^2{\bf 1_N}$.
In order to avoid the Goldstone bosons, we explicitly break the global symmetry by splitting the fields as
 \be
 \label{fluc}
 \Phi=\pmatrix{Y_{IJ} & Z_{Ia}\cr \tilde Z_{aI}&\hat \Phi_{ab}} \ ,\quad q=\pmatrix{\chi_{IJ}\cr \rho_{Ia}} \ ,\quad \tilde q^t=\pmatrix{\tilde \chi_{IJ}\cr\tilde\rho_{aI}} \ ,
 \ee
where $I,J=1,\ldots,N_f-N_c\equiv N$ and $a,b=1,\ldots,N_f-N=N_c$ and split the linear term
\be
- h\mu^2\Tr\Phi  \rightarrow -hm^2\Tr\,Y-h\mu^2\Tr\,\hat \Phi \ .
\ee
We will see that we need to work in the regime of parameters $\mu < m$.
This corresponds in the electric theory to having $N_c$ light flavours $(\tilde{Q}_a, Q_a)$ and $N_f-N_c$ heavier ones
$(\tilde{Q}_I, Q_I)$.\footnote{It has been shown in \cite{Giveon:2008wp} that $SU(N_c)$ SQCD with a number of light flavors
less than $N_c$ does not have an ISS metastable vacuum, due to a two loop effect that destabilizes it. In our case the number of light flavors in the electric description is $N_c$ and this two loop effect is absent.} Next, we need to break the R-symmetry, which we will do explicitly by
adding  a mass term to the off diagonal components of the singlet $h^2m_z\Tr\,\tilde Z Z$ \cite{Kitano:2006xg}.
This corresponds to a quartic coupling of the electric quarks $Tr(Q_aQ_I\tilde{Q}_a \tilde{Q}_I)$.

The final superpotential reads
 \be\label{super}
 W=h\Tr\, \left(\tilde\chi Y\chi+\tilde \rho Z\chi+\tilde\chi\tilde Z\rho+\tilde\rho\hat\Phi\rho\right)-hm^2\Tr\,Y-h\mu^2\Tr\,\hat \Phi+h^2m_z\Tr\,\tilde Z Z \ .
 \ee
We will use this model for a direct mediation of supersymmetry breaking and analyze its phenomenological features.
The relevant parameters are the dimensionless coupling $h$, the dimension one mass parameters $(\mu,m,m_z)$ and the dimension one
magnetic scale $\Lambda_m$.
At energies $E < \Lambda_m$ we have the weakly coupled magnetic description (\ref{super}) with a canonical Kahler
potential and at $E > \Lambda_m$ we have an electric description. At certain higher energies we need UV completion.
We will constrain the parameters by consistency and experimental requirements.

The mass term $h^2m_z\Tr\,\tilde Z Z$ breaks R-symmetry, thus allowing gaugino masses. Generically, such a breaking creates new supersymmetric vacua \cite{Nelson:1993nf}, and the longevity
of the metastable vacuum requires small gaugino masses compared to the scalar masses (split supersymmetry).
However, the superpotential (\ref{super}) is not generic, i.e.
a generic one would include also quadratic and cubic terms in $\hat\Phi$,
and it does not introduce new supersymmetric vacua.
We will see that
dimensionless parameter $\frac{m_z}{m}$ controls the split between the gaugino and squarks masses, while
the longevity of the metastable vacuum is controlled by $\frac{\mu}{m}$.
Having two parameters will allow us to avoid split susperymmetry, while maintaining longevity.

\subsection{Classical vacua}

The model (\ref{super}) does not have classical supersymmetric vacua, but only supersymmetry breaking ones.
Nonperturbatively, a supersymmetric  vacuum appears, parametrically far in field space \cite{Kitano:2006xg}.

The classical vacua are:
\begin{itemize}

\item

The ISS vacuum :
 \be\label{issvev}
 \chi_{IJ}=m\,\delta_{IJ},\quad \tilde\chi_{IJ}=m\,\delta_{IJ} \ ,
 \ee
and all other fields in (\ref{super}) have zero vev. $\hat\Phi$ is a pseudomodulus, namely it is a classically flat direction which is lifted at one loop. This is the vacuum on which we will base
the analysis. The superpotential around this vacuum takes the form
 \bea\label{superiss}
 W=h\Tr\, \left(\tilde \rho Z\chi+\tilde\chi\tilde Z\rho+\tilde\rho\hat\Phi\rho+ m\tilde\rho Z+m\tilde Z\rho
 -\mu^2\,\hat \Phi+h m_z\,\tilde Z Z\right) +\ldots\
 \eea
where we shifted $\chi\to m+\chi$ and $\tilde\chi\to m+\tilde\chi$ and we omitted the terms involving $Y$, which are not relevant for the rest of the discussion. The classical vacuum energy is
\be
V = V_{ISS}=(N_f-N)\vert h\mu^2\vert^2 \ .
\ee

At one loop, a potential for the pseudomodulus is generated that gives a mass and an expectation value to $\hat\Phi$
 \be\label{oneloop}
 V^{1}(\hat\Phi)=M_{\hat\Phi}^2\Tr_{N_f-N}\vert \hat\Phi-\hat\Phi_0\vert^2 \ ,
 \ee
where the explicit values of $M_{\hat\Phi}$ and $\hat\Phi_0$ are given by (\ref{oneloopo}) and (\ref{effe}).

\item

$N$ additional  supersymmetry breaking vacua where, on top of (\ref{issvev}), we have also
 \bea\label{othervev}
 \rho=\tilde\rho^t=&\mu\,\Id_n\ ,\cr
 Z^t=\tilde Z=&-{\mu m\over hm_z}\Id_n\ ,\cr
 Y={\mu^2\over hm_z}\Id_N \ ,&\quad \hat\Phi={m^2\over hm_z}\Id_n \ ,
 \eea
with classical vacuum energy
\be
V_n=(N_f-N-n)\vert h\mu^2\vert^2, \,\,\,\,\,\,\, n=1,\ldots,N \ .
\ee

Thus, the lowest energy perturbative supersymmetry breaking vacuum of the theory is given by $n=N$.

\end{itemize}

Nonperturbatively, a dynamical superpotential is generated, which introduces supersymmetric vacua related to gaugino condensation in the $SU(N)$ gauge group. These extra vacua are very far in the $\hat \Phi$ field direction \cite{Kitano:2006xg}.

\subsection{R-symmetry breaking}

R-symmetry breaking generates soft masses for the gauginos.
In the limit of vanishing $m_z$, the model reduces to the original ISS one \cite{Intriligator:2006dd}, which has an R-symmetry with
\be
R(\Phi)=2,~~~~~~~R(q)=R(\tilde q)=0 \ ,
\ee
that
forbids gaugino masses but not sfermion masses, when we embed the MSSM gauge group into the flavor group.  There is a tension between R-symmetry breaking which generically introduces new supersymmetry vacua, raising the issue of longevity of the vacuum and having gaugino and sfermion masses of the same order, leading to a split supersymmetry scenario.

Our model is in the regime $\mu/m<<1$ and $m_z/m \sim 1$.
When $\mu=0$, the theory does not reduce to the ISS one: the F-term
$F^\dagger_{\hat\Phi}$ vanishes, restoring supersymmetry.
The moduli space of supersymmetric vacua is parameterized by $\hat \Phi$ and Nelson-Seiberg's theorem \cite{Nelson:1993nf} is evaded, because in the vacuum (\ref{issvev}) there is a {\it different} (than the ISS one) unbroken R-symmetry $U(1)_{R'}$, under which $\hat \Phi$ has zero R-charge and
\be
R'(\rho)=R'(\tilde\rho)=R'(Z)=R'(\tilde Z)=1, \quad R'(Y)=2,~~~~~~R'(\hat \Phi)=0 \ .
\ee
Embedding the MSSM gauge group into the global $SU(N_f-N)$, and parameterizing the gaugino and scalar masses schematically as
\be
\Lambda_g=F_{\hat\Phi}\times R_\half,~~~~~~
\Lambda_s^2=\vert F_{\hat\Phi}\vert^2R_0^2 \ ,
 \ee
 then $U(1)_{R'}$ allows for $R_\half$ and $R_0$ of the same order, but it restores supersymmetry enforcing $F_{\hat\Phi}=0$.
 Note, in comparison, that the R-symmetry of the original ISS model is problematic for phenomenology because it enforces $R_\half=0$, with $R_0$ and $F_{\hat\Phi}$
non-vanishing.
When we switch on a small $\mu$, we break
the $U(1)_{R'}$ explicitly and supersymmetry spontaneously, by the vev of $F_{\hat\Phi}$. Moreover, we do not introduce any new supersymmetry vacua coming in from infinity of field space, as it would happen if the superpotential deformation that breaks explicitly R-symmetry were generic in the sense of \cite{Nelson:1993nf}.

\section{Direct mediation of supersymmetry breaking}
In order to build a model for direct mediation of supersymmetry breaking, we embed the Standard Model gauge group in the global symmetry
group $SU(N) \times SU(N_f-N)$, i.e. we get the Standard Model gauge group by gauging a subgroup of the flavor symmetry group.
The embedding of the MSSM into $SU(N)$ has been discussed by \cite{Kitano:2006xg}. In that case, one achieves perturbative unification but the gravitino mass exceeds the cosmological bounds \cite{Viel:2005qj}. It might be possible to get in this embedding a light gravitino, though giving up perturbative unification. There is also a Goldstone boson for the broken ISS baryon symmetry that is charged under the MSSM gauge groups; this would be problematic for cosmology, but it can be given a mass by gauging the baryon symmetry.

We follow a different route and embed the MSSM gauge group in the unbroken flavor symmetry group $SU(N_f-N)$ and require $N_f-N\geq 5$.
In the analysis we will take $N_f=6,\, N=1$, so the DSB sector reduces to a deformation of an O'Raifeartaigh model, and we will use the metastable vacuum (\ref{issvev}).
The messenger fields are $\{\rho,\tilde\rho,Z,\tilde Z\}$, and they couple through the superpotential to ${\hat \Phi}$ whose F-term $F_{\hat \Phi}$ breaks supersymmetry.
Note that $\{\rho,\tilde\rho,Z,\tilde Z\}$ and ${\hat \Phi}$ are charged under the MSSM gauge
group and couple to the MSSM fields through gauge interactions. For the reader's convenience, we collected their MSSM quantum numbers in Appendix C.

The important scales in gauge mediation models are the messengers mass and the supersymmetry breaking scale. Their ratio times a gauge loop factor determines the scale of soft supersymmetry breaking terms. This model has two additional scales: $\Lambda_m$, which is the cutoff of the magnetic theory, and the mass of the pseudomodulus $\hat\Phi$. This mass is generated by the Coleman-Weinberg potential, and similarly to the soft mass terms it is determined by the ratio of the supersymmetry breaking scale and the messenger masses. However, instead of the gauge coupling, we multiply by the DSB sector Yukawa coupling $h^2$ which leads to a new scale in the theory. The various scales and their dependence on the input parameters are depicted in figure \ref{fig:ScalesPlot}. Note also that in this model the messengers are linear combinations of $\rho$ and $Z$, therefore we have two messengers with different masses $m_{\frac{1}{2}-}$ and $m_{\frac{1}{2}+}$.
\FIGURE[t]{\includegraphics[width=9.5cm,height=9.5cm]{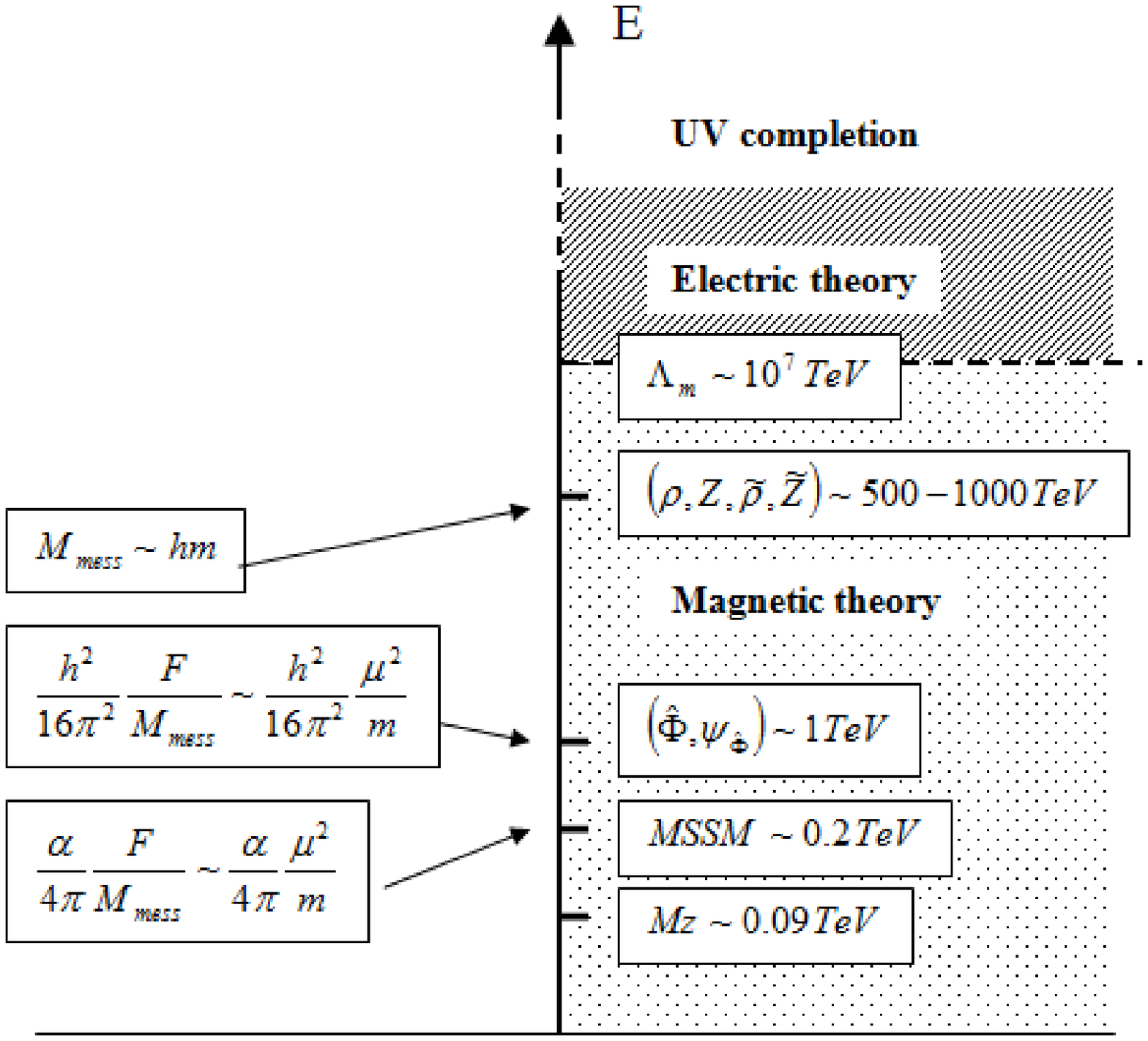}
\caption{The various energy scales and parameters of the model.\label{fig:ScalesPlot}}}
In the following we will present the constraints on the parameters of the model and the features it presents. Details of the spectrum and predictions of the model will be given later. Aspects of the analysis are outlined in the appendices.

\subsection{Constraints on the parameter space}
\label{constraintson}

The direct mediation model contains one dimensionless coupling $h$, three dimensionful parameters $(\mu,m_z,m)$ and the magnetic scale $\Lambda_m$. We now briefly list the phenomenological and theoretical constraints imposed on the parameter space:

\begin{itemize}

\item{$h$}:
The dimensionless parameter $h$ can be written up to an order one constant in terms of the magnetic and electric scales.
$\frac{h}{4\pi}$ is used for a perturbative expansion, therefore we require that $h$ is at most $\sim O(1)$. When we analyze
in detail the spectrum of the model and take into account the LEP bound on the Higgs mass we find that $h>1$. We will present
a detailed analysis for the case $h=2$.

\item{$h$,$\mu$}:
The gravitino has to be light in order to be consistent with cosmological bounds \cite{Viel:2005qj}, i.e.
\be\label{graviti}
 m_{3\over2}={F\over \sqrt{3}M_{Pl}} < 16 eV \ .
 \ee
where the supersymmetry breaking scale $F$ is the square root of the value of the potential at the supersymmetry breaking minimum. This can be translated into a constraint on $h$ and $\mu$:
 \be
h\mu^2 = \frac{F}{N_f-N} < (150 {\rm TeV})^2 \ .
 \ee

\item{$m$,$\mu$}: The relation between the parameters $\mu$ and $m$ has two effects. On the one hand
the ratio $\frac{\mu}{m}$ controls the longevity of the metastable vacuum and we get an upper bound $\frac{\mu}{m}<\frac{1}{5}$. On the other,
the ratio $\frac{\mu^2}{m}$ determines the soft supersymmetry breaking terms ($m$ controls the messenger masses while $\mu$ controls the supersymmetry breaking scale) and is therefore constrained from below by bounds from the MSSM spectrum.

\item{$m_z$}: The parameter $m_z$ plays a triple role. It controls the R-symmetry breaking, allowing gaugino masses.
The dimensionless parameter $\frac{m_z}{m}$ controls the split between the gaugino and squarks masses, i.e. for
$\frac{m_z}{m} \sim 1$ we avoid split supersymmetry.

In order to avoid a negative mass for the messenger we require
\be
\vert m^2\pm hm_z\hat\Phi_0\vert^2>\mu^2(m^2+h^2m_z^2) \ ,
\ee
this can be translated to constraints on $m_z$.
The parameter $m_z$ also plays a role in determining the lifetime of the metastable vacuum, leading to an upper bound. This bound depends on the value of $h$.

\item{$\Lambda_m$}: The scale $\Lambda_m$ is the scale at which the weakly coupled magnetic description (\ref{super}) breaks down: at energies $E > \Lambda_m$ we have an electric description. Nonpertubative effects restore supersymmetry at large values of $\hat \Phi$ \cite{Intriligator:2006dd}. To suppress the decay to this true vacuum it is sufficient that $\Lambda_m/m>5$. On the other hand, requiring that the full messenger spectrum lies below the cutoff scale, we need approximately $\Lambda_m/m>10$. We will postpone the discussion of the UV completion of the model to section \ref{cascading}.

\end{itemize}
We are thus lead to a relatively small range in parameter space. Compiling all the above considerations leads to a representative
set of  values for the input parameters:
 \bea\label{ConstrainedParameters}\quad &h\sim 2,\quad \mu\sim 100\, {\rm TeV},  \quad m\sim 500{\,\rm TeV}, \quad\Lambda_m> 5000 {\,\rm TeV}
, \cr
&m_z<220\tev\quad {\rm or}\quad 330\tev<m_z<650\tev  \ .
 \eea
Other values of $h$ in its allowed range will also lead to reasonable phenomenology with similar features.

Let us give some more details on the constraints mentioned above.

\subsection{Longevity}
\DOUBLEFIGURE[t]
{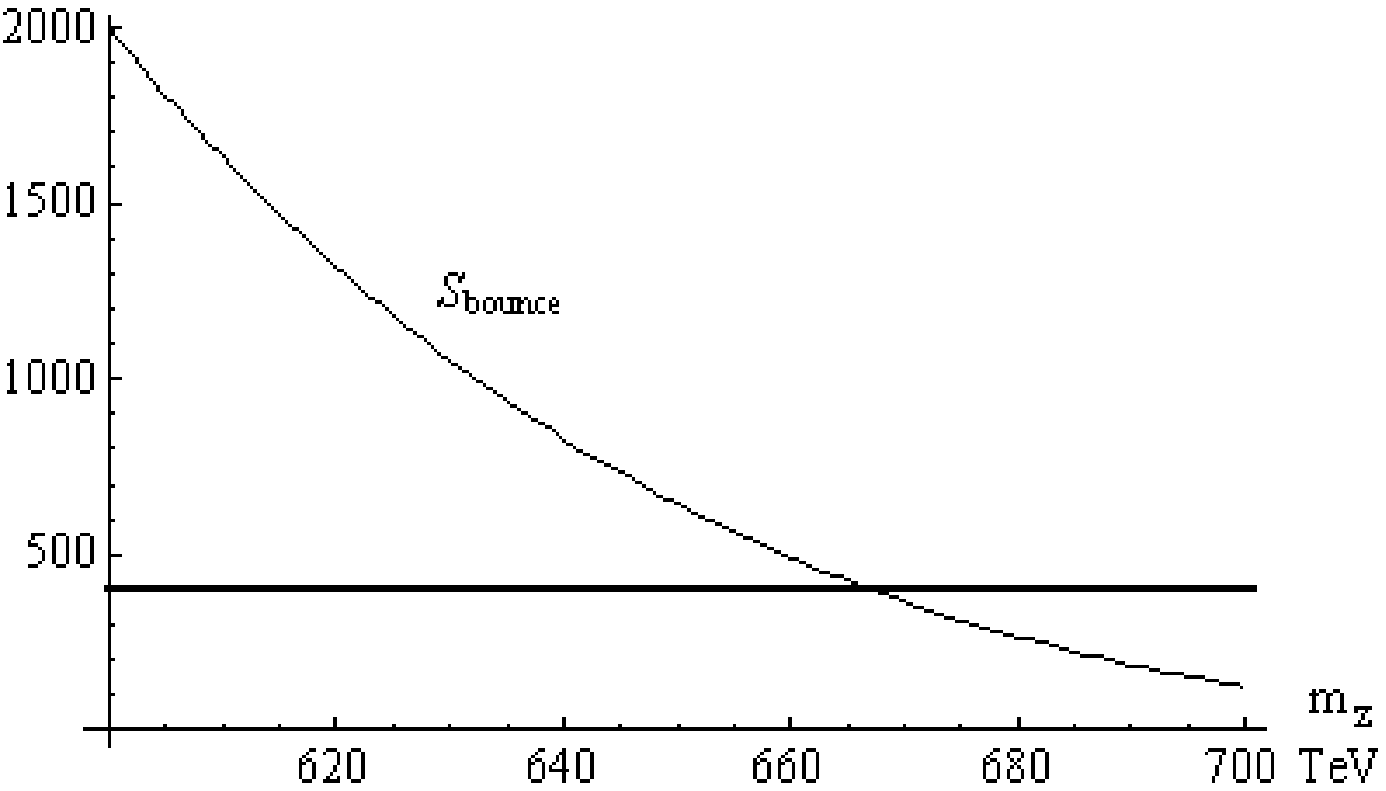,width=7.5cm,height=5cm}{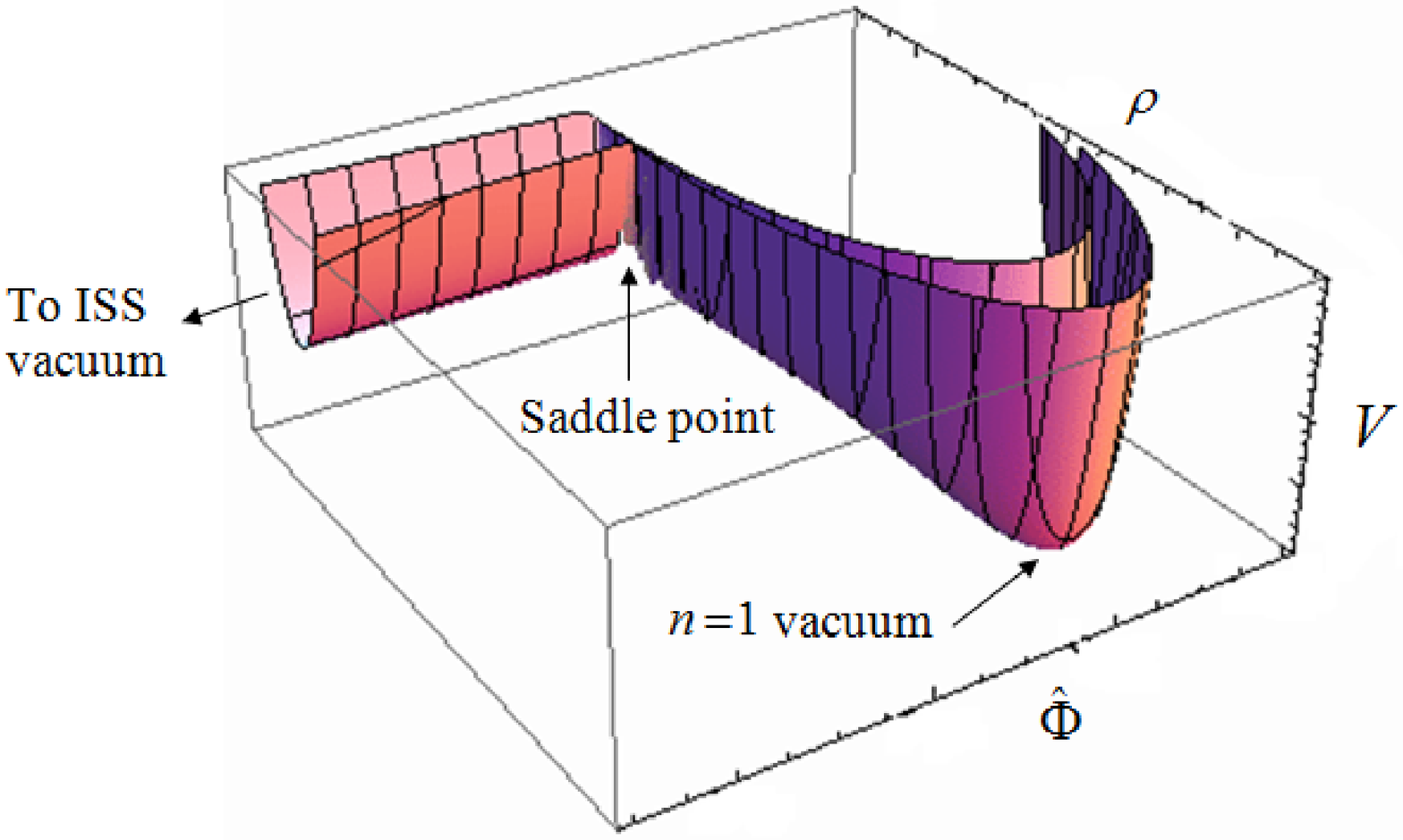,width=9cm,height=5.5cm}{Numerical evaluation of the bounce action for the decay to the closest vacuum in figure \label{sbounce} as a function of $m_Z$.\label{bounceaction}}{The effective potential for a real slice of the potential $V_{eff}(\hat\Phi)$ for the $\hat\Phi$ bounce trajectory. The plot is evaluated at $m_Z=150\tev$.
\label{potential3D}}

The ISS vacuum can decay either to the closest metastable vacua (\ref{othervev}) or to the supersymmetric vacuum generated by nonperturbative effects (in the case the magnetic gauge group is non-empty supersymmetry restoration is related to gaugino condensation). In order to have a long lifetime we require that the euclidean bounce action $S_{bounce}$ for the decay from the ISS into another vacuum is
 \be\label{bounceco}
 S_{bounce}>400 \ .
 \ee

In appendix B we evaluate the decay probability per unit time and unit volume from the ISS vacuum to the closest supersymmetry breaking vacuum (\ref{othervev}), for $n=1$.
In figure \ref{sbounce} we plot the bounce action and in figure \ref{potential3D}
the effective potential for the bounce trajectory.

Let us consider now the nonperturbative supersymmetric vacuum. It is very far in field space along the $\Phi$ direction, its position being proportional to the magnetic dynamical scale $\Lambda_m$. The decay to this vacuum has been evaluated in \cite{Kitano:2006xg} using the triangle approximation. The euclidean bounce action is approximately given by
\be\label{bouncesusy}
 S_{bounce}\sim \left({m\over\mu}\right)^4\left({\Lambda_m\over m}\right)^{4(N_f-3N)\over N_f-N} \ .
 \ee
The decay is approximately independent of $m_z$ and by using our parameters we find that the bounce action is much larger than the requirement (\ref{bounceco}).

\subsection{Gaugino and squarks masses}

\label{sec:Gaugino_and_squarks_masses}

We work in a regime where the F term $F_{\hat\Phi}$ is smaller than the messenger scale
$\mu^2/hm^2<<1$,
 and can use simple expressions to compute the gaugino and scalar soft masses. The gaugino masses are
 \bea\label{gaugino}
 m_r=&{\alpha_r\over4\pi} F_{\hat\Phi}\partial_{\hat\Phi}\det\log{\cal M} \cr
 =&{\alpha_r\over4\pi} F_{\hat\Phi}\sum_{\pm}
 {\partial_{\hat\Phi}{\cal M_\pm}\over {\cal M_\pm} }\ ,
 \eea

where ${\cal M}$ is the superpotential mass matrix\footnote{We assume that doublet and triplet messengers have the same mass. In this case, the dangerous negative contribution to the sfermion masses, proportional to the hypercharge D-terms, are absent \cite{Dimopoulos:1996ig}.}
 \be\label{massma}
 {\cal M}=\pmatrix{h\hat\Phi_0&hm\cr hm&h^2m_z} \ ,
 \ee
and ${\cal M}_\pm$ its eigenvalues
\be\label{eigen}
{\cal M}_\pm=\left\vert\half h \left(h m_z+\hat\Phi_0 \pm\sqrt{4 m^2+(-h m_z+\hat\Phi_0 )^2}\right)\right\vert \ ,
\ee
and $\Phi_0$ is given in the appendix (\ref{oneloopo}).
The final expression reads
 \bea\label{gaugi}
 m_r=&{\alpha_r\over4\pi}\Lambda_g \ ,\cr
 \Lambda_g=&N{h^2\mu^2m_z\over m^2-h\hat\Phi_0 m_z}\ .
 \eea
The scalar masses are given by
 \bea\label{scala}
 m_{\tilde f}^2=&\sum_{r=1}^3 2C_{\tilde f}^r\left({\alpha_r\over4\pi}\right)^2\Lambda_s^2 \ ,\cr
 \Lambda_s^2=&\half N\vert F_{\hat\Phi}\vert^2{\partial^2\over \partial\hat\Phi\partial\hat\Phi^\dagger}\sum_{\pm}\left(
 \log\vert {\cal M}_\pm\vert^2\right)^2 \ ,\cr
 =&N\vert F_{\hat\Phi}\vert^2\sum_{\pm}\left\vert {\partial_{\hat\Phi}{\cal M}_\pm\over {\cal M}_\pm}\right\vert^2 \ .
 \eea

The gaugino masses and the scalar masses share the same dependence on the small parameter $\mu/m$, which is the one that controls the longevity of the vacua and the breaking of supersymmetry. Hence, in the vacuum (\ref{issvev}), we can relax the tension between having a long lived metastable vacuum and large gaugino masses, thus avoiding a split supersymmetry spectrum.
 \FIGURE{\includegraphics[width=7cm,height=4cm]{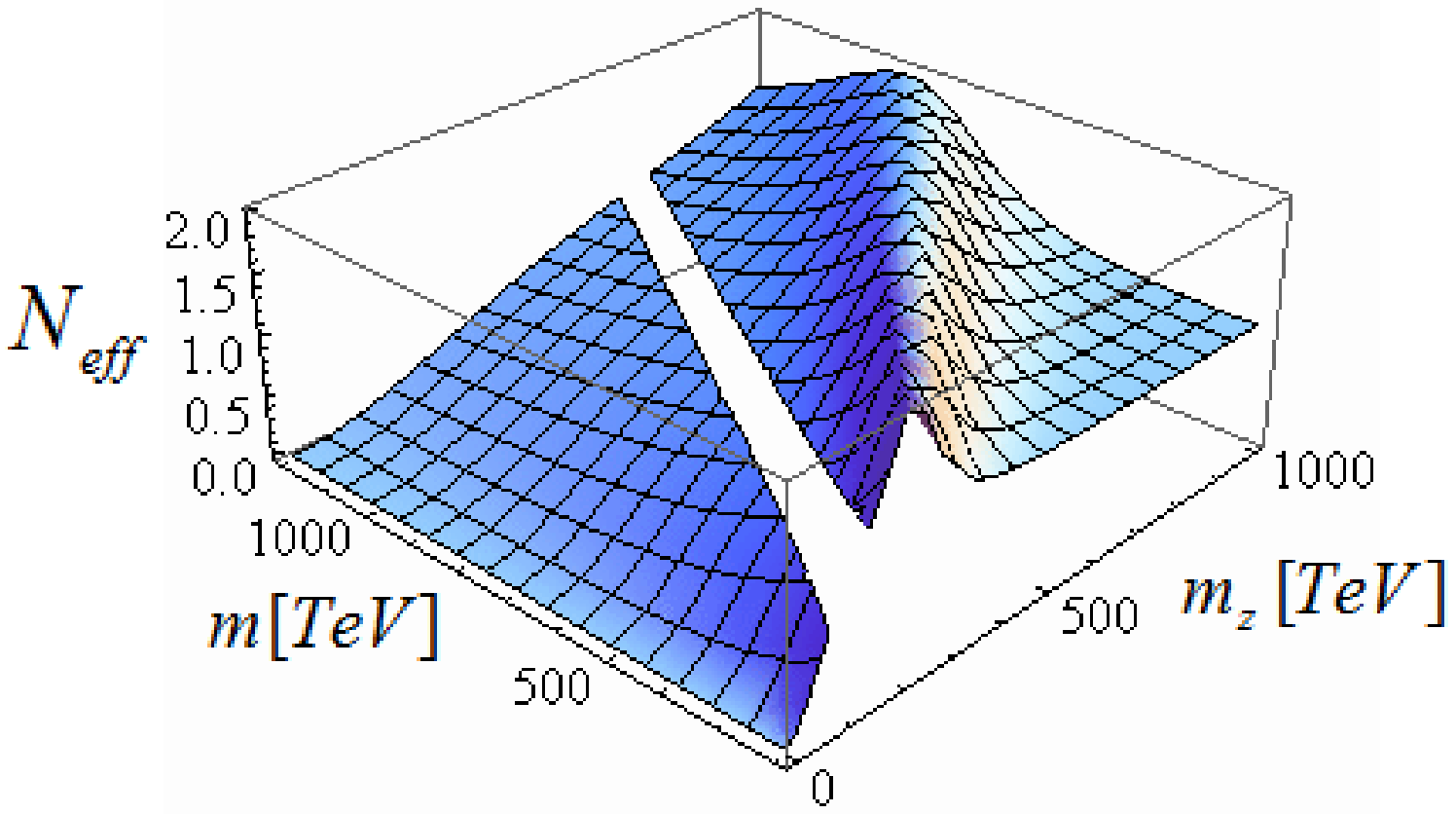}
\caption{The effective number of messengers as a function of $\frac{m_z}{m}$ varies between zero and two. The plot is disconnected in the
regime where $m_z$ is not allowed.
\label{fig:Neff3D}}}

 In ordinary gauge mediation models the number of messengers is the ratio $N_{mess}=\Lambda_g^2/\Lambda_s^2$.
 This is not the case in our model, and we can define an effective messenger number \cite{Cheung:2007es}
 \be\label{Neff}
 N_{eff}(m_z)=\Lambda_g^2/\Lambda_s^2 \ .
 \ee
By varying continuously $\frac{m_z}{m}$, inside the region allowed by the phenomenological constraints,
 $N_{eff}$ varies between zero and two (the number of messengers) as is shown in figure
\ref{fig:Neff3D}.

As mentioned above, the LSP in the model is the gravitino. A decay of an NLSP $\tilde{\chi}$ to the LSP gravitino and a Standard Model particle, $\tilde{\chi} \rightarrow
SM + \tilde{G}$ is characterized by a decay rate $\Gamma \sim \frac{m_{\tilde{\chi}}^5}{16 \pi F^2}$, yielding
a life time $\tau \sim 10^{-12} sec$.

\section{A visible supersymmetry breaking sector}

A crucial prediction of our model of direct mediation is the presence of light particles coming from the supersymmetry breaking sector (figure \ref{fig:PhiMasses}).
They are the fluctuations of some pseudomoduli and their superpartners, whose mass only arises at one loop and hence it is suppressed by a $16\pi^2$ factor with respect to the typical scale of the DSB sector, the messenger mass. Choosing a messenger mass of a few hundred TeV one obtains therefore some exotic particles of a few TeV or lower. By embedding the MSSM gauge group into the DSB sector unbroken flavor symmetry, we give MSSM quantum numbers to these light DSB sector particles. In our model, the pseudomodulus comes from the traceless part of the chiral superfield $\hat\Phi$, in the adjoint representation of $SU(5)$ (its trace part is the Goldstino), which decomposes in the following way under $SU(3)\times SU(2)_L \times U(1)_Y$
 \be
 \label{decompose}
 {\bf 24}=({\bf 8},{\bf 1})_0\oplus({\bf 1},{\bf 3})_0\oplus
 ({\bf 3},{\bf2})_{-5/6}\oplus({\bf \bar 3},{\bf2})_{5/6}\oplus ({\bf 1},{\bf 1})_0 \ ,
 \ee
and we split accordingly the bosonic and fermionic components of $\hat\Phi$ as
 \bea\label{visiphi}
 \hat\Phi=&\varphi_8\oplus\varphi_3\oplus\tilde p\oplus\tilde p' \oplus S \ , \cr
 \psih=&\Psi_8\oplus \Psi_3\oplus\Psi_{\tilde p}\oplus \Psi'_{\tilde p}\oplus\Psi_S \ ,
 \eea
Note in particular the presence of a singlet fermion $\Psi_S$, that will play the role of lightest DSB sector particle (LHP) in some regions of the parameter space.
The boundary conditions for the mass of (\ref{visiphi}) are the one loop values $M_{\hat\Phi}$, common for all the scalars (\ref{oneloopo}), and $M_{\psih}$, common for all the fermions (\ref{diagram}).\footnote{We neglected the gauge contribution to their soft masses, which differs according to the quantum numbers. Since it is of order the MSSM soft masses, it is significantly smaller than the leading DSB Yukawa contribution.}
Starting from this value at the messenger mass, we run all the way down to the TeV scale, coupling their RG flow equations to the MSSM ones as explained in appendix \ref{rgfloww}. We computed their RG improved masses by a modification of the SoftSUSY algorithm.
After the RG evolution, the masses of such particles will split according to the usual pattern: colored particles become heavier than weakly interacting ones.
We will discuss this light sector in some detail.

\subsection{Singlet}
\label{singlesec}

Consider the singlet fermion $\Psi_S$. Its decay channels are two: a coupling to the Goldstino and a Yukawa interaction with the messengers. In the theory MSSM plus DSB sector, there is an exact R-parity, that combined with a second approximate $Z_2$ symmetry will strongly suppress the decays of the LHP. The exact R-parity, which we will denote by R, is the usual R-parity of the MSSM combined with the DSB sector R-parity, under which the DSB sector bosons (fermions) are even (odd). The second approximate $Z_2$ symmetry, that we will call $P$, is the usual R-parity of the MSSM with a DSB sector parity under which the bosons are odd and the fermions are even. This discrete symmetry is explicitly broken by the Yukawa interaction (\ref{superiss}), so the pseudomodulus can only decay through a loop of the messengers.

\FIGURE[t]{
\begin{picture}(200,76) (10,-19)
   \SetWidth{0.5}
   \SetColor{Black}
   \Text(20,11)[lb]{\Large{\Black{$S$}}}
   \DashLine(50,16)(90,16){8}
   \ArrowLine(90,16)(135,-14)
   \ArrowLine(135,46)(90,16)
   \ArrowLine(180,46)(135,46)
   \ArrowLine(135,-14)(180,-14)
  \Text(90,-14)[lb]{\Large{\Black{$\Psi_\rho$}}}
   \Text(90,31)[lb]{\Large{\Black{$\Psi_{\tilde{\rho}}$}}}
   \Text(190,-19)[lb]{\Large{\Black{$\tilde N$}}}
   \Text(190,41)[lb]{\Large{\Black{$\tilde N$}}}
   \DashLine(135,-14)(135,46){8}
   \Gluon(135,46)(180,46){5}{4}
   \Gluon(135,-14)(180,-14){5}{4}
 \end{picture}
\caption{The effective vertices for the decay of a scalar into 2 neutralinos.}
\label{fig_singlet_decay}}

In the effective theory below the messenger scale, the leading gauge invariant $C$-parity conserving decay of the scalar $S$ into MSSM is via two neutralinos, as described in figure in figure \ref{fig_singlet_decay}.


The leading order decay channel of the LHP $\Psi_S$ will proceed through
an emission of a gravitino and a similar decay into two neutralinos
 \be
 \label{leading}
 \Psi_S\rightarrow \tilde N\tilde N\tilde G \ ,
 \ee
Because this decay is hugely suppressed by a loop factor and (DSB sector) GIM mechanisms, the singlet is possibly long-lived. Depending on the details of its lifetime and its relic abundance, it may provide a suitable dark matter candidate in some region of our parameter space.

\subsection{Colored}

\FIGURE[t]{\includegraphics[width=9cm,height=6.5cm]{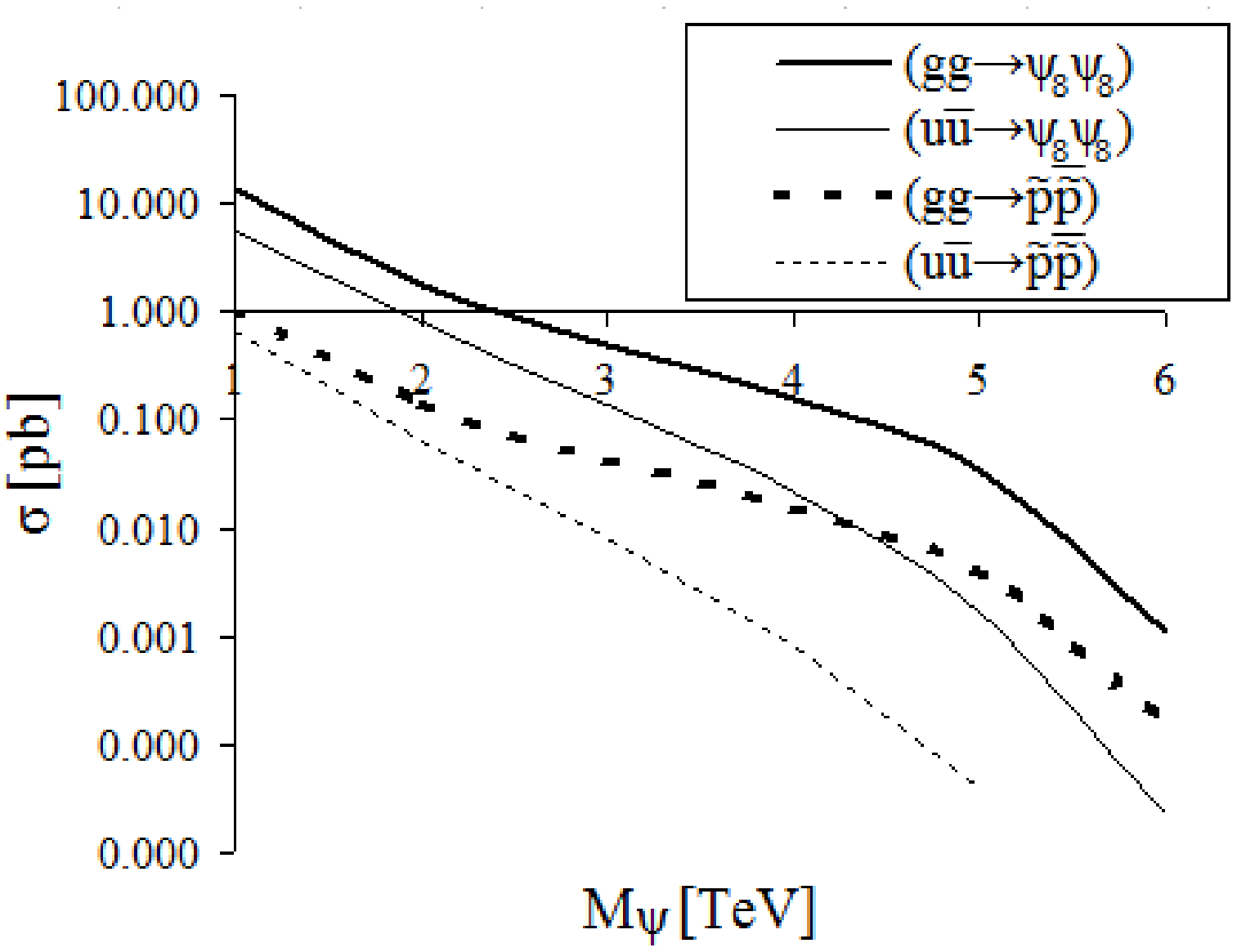}
\caption{The cross section for production of supersymmetry breaking sector colored particles in the LHC as a function of their mass\label{PsiCS}}}

Let us estimate the production at LHC of the light DSB sector particles. The colored particles $({\bf 8},{\bf 1})_0$ and $({\bf 3},{\bf 2})_{-5/6}$, $({\bf \bar3},{\bf 2})_{5/6}$ are most likely produced, while the weakly interacting ones and the singlet will have a much slower rate. As an example, let us estimate the production of the $({\bf 3},{\bf 2})_{-5/6}$ scalars, that we denoted by $\tilde p$. The electroweak doublet consists of two squarks with electric charges $Q_{\rm em}=(-1/3,-4/3)$. At the LHC, the production of such particles occurs through the scattering of two gluons and of a quark-antiquark pair. In the leading parton approximation, we can adapt the cross section for the production of squarks \cite{Beenakker:1996ch} (setting to zero the Yukawa coupling contribution)
 \bea
 \label{partoncross1}
 \sigma_{gg\to\tilde p\bar{\tilde p}}(s)&=&{2\pi\alpha_s^2\over s}\left[
 \beta_{\tilde p}\left({5\over24}+{31 m_{\tilde p}^2\over12s}\right)+\left({4m_{\tilde p}^2\over 3s}+{m_{\tilde p}^2\over 3s^2}\right)\log\left({1-\beta_{\tilde p}\over 1+\beta_{\tilde p}}\right)\right] \nonumber \\
 \sigma_{q_i\bar q_j\to \tilde p\bar{\tilde p}}(s)&=&{\delta_{ij}}{2\pi\alpha_s^2\over s}\beta_{\tilde p}\left({4\over27}-{16 m_{\tilde p}^2\over27 s}\right) \ .
  \eea
where $\beta_{\tilde p}=\sqrt{1-4m_{\tilde p}^2/s}$ and $m_{\tilde p}$ is the mass of $\tilde p$ as we RG evolved it down to the LHC energies.\footnote{Since the final states are $SU(2)$ doublets we include an overall factor of two in (\ref{partoncross1}).} The total hadronic cross section for the production of ${\tilde p}$ through proton-proton$\to \tilde p{\bar{\tilde p}}$ scattering is the convolution of (\ref{partoncross1}) with the parton distributions $f_i(x)$\cite{Tung:2006tb} in the proton at leading order
 \be
 \label{sigmatotal}
 \sigma_{\tilde p\bar{\tilde p}}(S)=\sum_{i,j=g,q,\bar q}\int\,dx_1\int\,dx_2f_i(x_1)
 f_j(x_2)\sigma_{ij}(s=x_1x_2S) \ ,
 \ee
where $\sqrt{S}=14$ TeV. In figure \ref{PsiCS} we plot the cross sections as a function of the RG evolved mass of $\tilde p$. Because there is no mixing matrix between the particles $(\Psi_{\tilde p},\tilde p)$ and the MSSM quarks and squarks, the former are possibly long-lived. They will therefore hadronize and produce exotic mesons. We leave the investigation of this issue for the future.

The estimate of production cross section at LHC for the light colored octet fermion $\Psi_8$ can be given as well, using the parton cross sections
 \bea
 \label{partoncross2}
 \sigma_{gg\to\Psi_8\Psi_8}(s)&=&{\pi\alpha_s^2\over s}\left[
 \beta_\Psi\left(-3-{51 M_\Psi^2\over4s}\right)+\left(-{9\over4}-{9M_\Psi^2\over s}+{9M_\Psi^2\over s^2}\right)\log\left({1-\beta_\Psi\over 1+\beta_\Psi}\right)\right] \nonumber \\
 \sigma_{q\bar q\to\Psi_8\Psi_8}(s)&=&{\pi\alpha_s^2\over s}\beta_{\Psi_8}\left({8\over9}+{16 m_{\Psi_8}^2\over9 s}\right) \ .
  \eea
where $\beta_{\Psi_8}=\sqrt{1-4M_{\Psi_8}^2/s}$ and $M_{\Psi_8}$ is the mass of $\Psi_8$ as we RG evolved it down to the LHC energies. This octet is long-lived, because of the suppression mechanism in the decay. The leading decay channel of such particle is
$\Psi_8\to \tilde g g$, namely it decays into a gluon and a gluino, through the effective vertex in figure \ref{fig_adjoint_decay}.

\FIGURE[t]{
\begin{picture}(200,76) (10,-19)
   \SetWidth{0.5}
   \SetColor{Black}
   \ArrowLine(50,16)(90,16)
   \Text(20,11)[lb]{\Large{\Black{$\Psi_\phi$}}}
   \ArrowLine(90,16)(135,-14)
   \DashLine(90,16)(135,46){8}
   \ArrowLine(135,46)(180,46)
   \Text(90,-14)[lb]{\Large{\Black{$\Psi_\rho$}}}
   \Text(90,31)[lb]{\Large{\Black{$\tilde{\rho}$}}}
   \Text(190,-19)[lb]{\Large{\Black{$g$}}}
   \Text(190,41)[lb]{\Large{\Black{$\tilde g$}}}
   \ArrowLine(135,-14)(135,46)
   \Line(133,14)(137,18)\Line(133,18)(137,14)
   \Text(140,23)[lb]{\Large{\Black{$\Psi_{\tilde \rho}$}}}
   \Text(140,1)[lb]{\Large{\Black{$\Psi_\rho$}}}
   \Gluon(135,-14)(180,-14){5}{4}
   \Gluon(135,46)(180,46){5}{4}
 \end{picture}
\begin{picture}(210,76) (10,-19)
   \SetWidth{0.5}
   \SetColor{Black}
   \ArrowLine(50,16)(90,16)
   \Text(20,11)[lb]{\Large{\Black{$\Psi_\phi$}}}
   \Line(90,16)(135,-14)
   \DashLine(90,16)(135,46){8}
   \ArrowLine(135,-14)(180,-14)
   \Text(90,-14)[lb]{\Large{\Black{$\Psi_\rho$}}}
   \Text(90,31)[lb]{\Large{\Black{$\tilde{\rho}$}}}
   \Text(190,-19)[lb]{\Large{\Black{$\tilde g$}}}
   \Text(190,41)[lb]{\Large{\Black{$g$}}}
   \DashLine(135,46)(135,-14){8}
   \Line(133,14)(137,18)\Line(133,18)(137,14)
   \Text(140,26)[lb]{\Large{\Black{$\tilde \rho^\dagger$}}}
   \Text(140,1)[lb]{\Large{\Black{$\rho^\dagger$}}}
   \Gluon(135,-14)(180,-14){5}{4}
   \Gluon(135,46)(180,46){5}{4}
 \end{picture}
\caption{The effective vertices for the decay of $\Psi_\phi$ into a gaugino and a gauge boson. The other two diagrams are obtained by exchanging the labels $\rho\leftrightarrow\tilde \rho$ in the loop.}
\label{fig_adjoint_decay}}

\subsection{Weakly interacting}

Let us briefly discuss the light DSB sector particle $\Psi_3$ in the $({\bf 1},{\bf 3})_0$. After electroweak symmetry breaking, the triplet will split into $(\psi^+,\psi^-,\psi^0)$, where the superscript denotes the electric charge. One loop electroweak effects will split the mass $M_{c}$ of the two charged particles with respect to the mass $M_0$ of the  neutral one. The mass of $\psi^\pm$ gets contribution by charged and neutral current interactions
\be
\delta M_{c}={\alpha_2\over\pi}M_{c}(\sigma_W+\cos\theta_W^2\sigma_Z+\sin \theta_W^2\sigma_\gamma) \ ,
\ee
where $\alpha_2$ is the running $SU(2)_L$ coupling and the loop integral
 \be
 \sigma_I=\int_0^1\,dx \ln\left({x\Lambda^2\over (1-x)^2M_c^2+xm_i^2}\right) ,
 \ee
where $I=W,Z,\gamma$ labels the gauge boson masses and $\Lambda$ is a UV cutoff. The mass $M_0$ of $\psi^0$ gets corrections from the charged current interaction only, $\delta M_0=(\alpha_2/\pi)M_02\sigma_W$. The relative mass shift $\Delta M=\delta M_c-\delta M_0$ between the masses is UV finite and amounts to $\Delta M/M_c=3\times 10^{-4}$, that is around $0.3$ GeV. Hence, $\psi^0$ is the lightest particle in the triplet and it decays only through charged current interactions. Its leading decay channel is into a photon or a Z boson and a neutralino, or into a W boson and a chargino.

\subsection{Neutral sector}

The part of the DSB sector containing the chiral multiplets $\chi,\tilde \chi$ and $Y$ can be produced from SM particles only through a loop of the messengers, hence its production cross section is very suppressed and it will not be produced at LHC.

The fermionic partner of the Goldstone boson for the $U(1)$ baryon symmetry of the ISS, which is classically massless, gets a mass at one loop through the Yukawa interaction with the messengers, the mass being of the same order of the adjoint fermion masses and may provide a cold dark matter candidate. We leave this issue to a future investigation.

\section{The detailed MSSM spectrum}

The low energy spectrum of the theory was calculated using a modified version of SoftSUSY 2.0 \cite{AllanachSOFTSUSY}. The modifications allow introduction of multiple messenger scales, adjustment of the MSSM $\beta$ functions to include the contribution of the light fields in the supersymmetry breaking sector ($\hat{\Phi}$), and they also enable running of the $\hat{\Phi}$ masses.

As discussed above, the seemingly large parameter space of the model is restricted to a narrow window by theoretical and phenomenological constraints. We chose to focus on the following set of parameters:
\bea
h=2~,~~\mu=100\tev~,~~m=500\tev~, 0.2 < m_z/m < 1.2.
\eea
The remaining parameter in the theory, $\Lambda_m$, does not affect the low energy spectrum. In addition to the parameters of the supersymmetry breaking sector, there are two more degrees of freedom introduced by the EWSB sector in the MSSM, for which we took the following values:
\bea
5<\tan\beta<35~,~~sgn(\mu)=\pm1~.
\eea
\DOUBLEFIGURE[t]{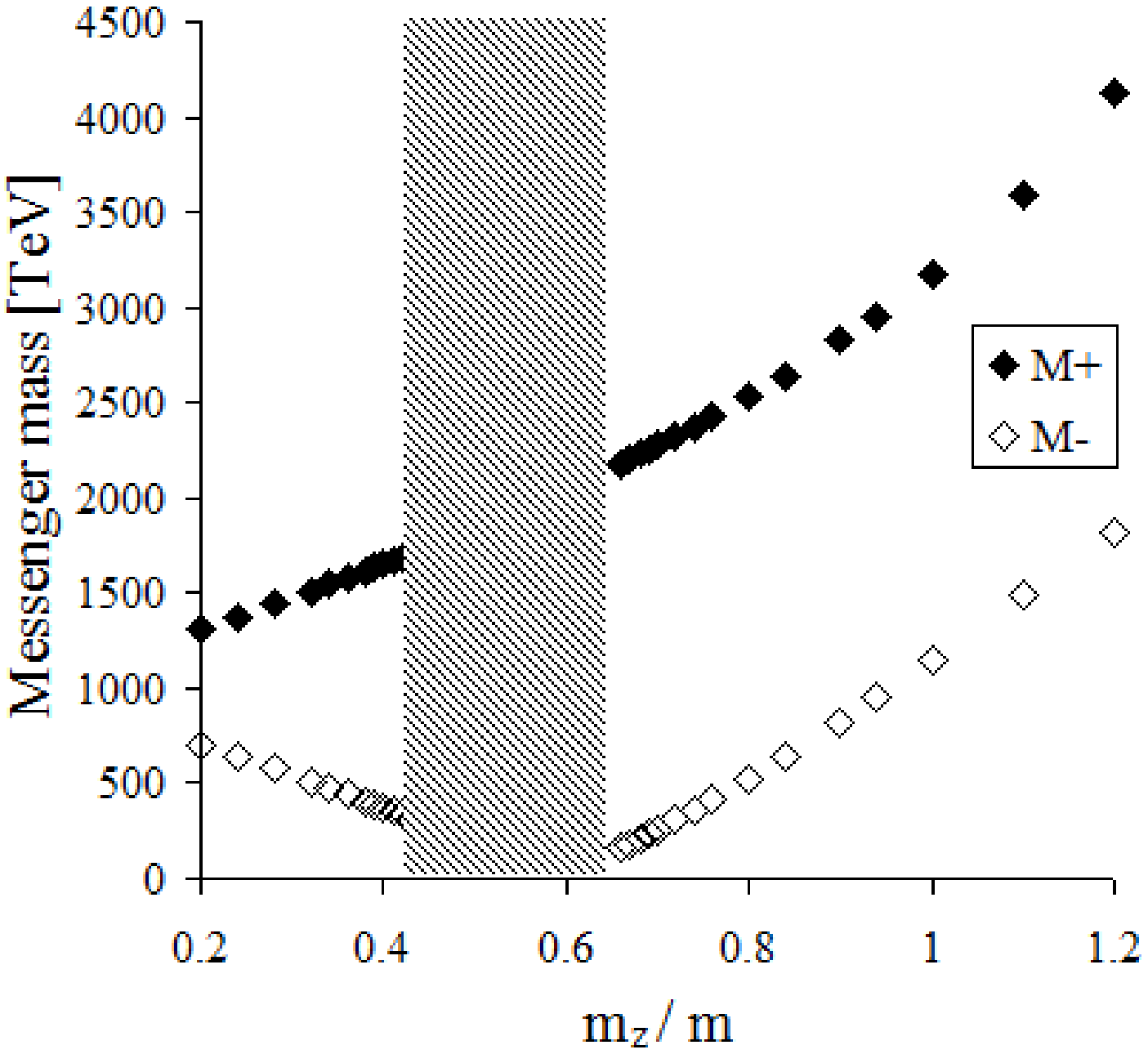,width=8cm,height=6cm}{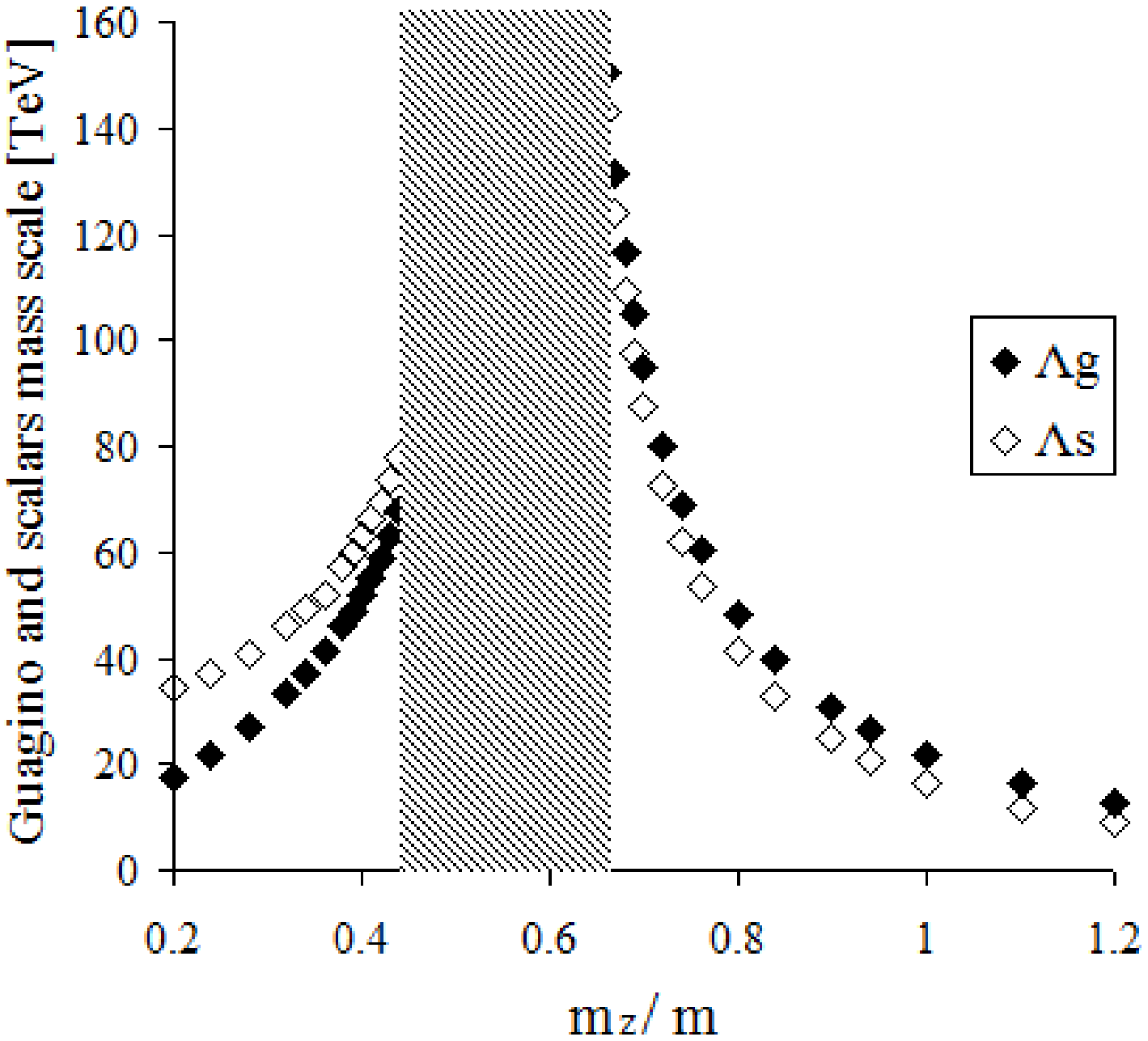,width=8cm,height=6cm}{The messenger mass as a function of $m_z/m$. In marked area the light messenger is too light or tachyonic, hence we exclude it.\label{fig:Messmass}}{The soft supersymmetry breaking mass scales, $\Lambda_s$ and $\Lambda_g$ as a function of $m_z/m$. Largest values are obtained close to the light messenger region.\label{fig:SoftMassScales}}

In order to understand the dependence of the spectrum on the parameter $m_z/m$, one should examine how it affects the messenger masses, and the gaugino and scalar mass scales ($\Lambda_g$ and $\Lambda_s$ - see plots \ref{fig:SoftMassScales} and \ref{fig:Messmass}). In the range $0.45 < mz/m < 0.65$ the light messenger becomes tachyonic, therefore this range is excluded. As one gets further away from this region, the messenger mass rises, leading to lower soft mass terms. Thus, the area of parameter space nearest to the tachyonic region leads to the highest soft mass terms. In the discussion below we show that this is an important condition for viable phenomenology.

The resulting spectrum has the general properties of ordinary gauge mediation with low supersymmetry breaking scale:
\begin{itemize}
\item{\textit{LSP}}: The LSP is a light gravitino ($<~16$ eV).
\item{\textit{NLSP}}: The NLSP is usualy a Bino like neutralino (20--200 GeV). For large $\tan \beta$ the NLSP can be a stau (see figure \ref{fig:SleptonMasses}).
\item There exists a hierarchy between colored and color singlet particles.
\end{itemize}
The new features that this model presents are:
\begin{itemize}
\item{\textit{Visible supersymmetry breaking sector}}: A new set of particles, charged under the SM gauge group, with masses in the range $1-10$ TeV. While the mass of the bosons does not vary much, the fermion masses are highly dependent on the ratio $m_z/m$, and they are split by the contribution of SM gauge loops to the RG flow. The lightest particle is thus either the fermionic singlet adjoint or one of the bosons. For $h=2$ the lowest mass they can get in the allowed range is $\sim 1$ TeV (plot \ref{fig:PhiMasses}). However, for lower values of $h$ one gets lower masses. This is in fact the feature of the model which is most influenced by the value of $h$.
\FIGURE[t]{\includegraphics[width=10cm,height=8.5cm]{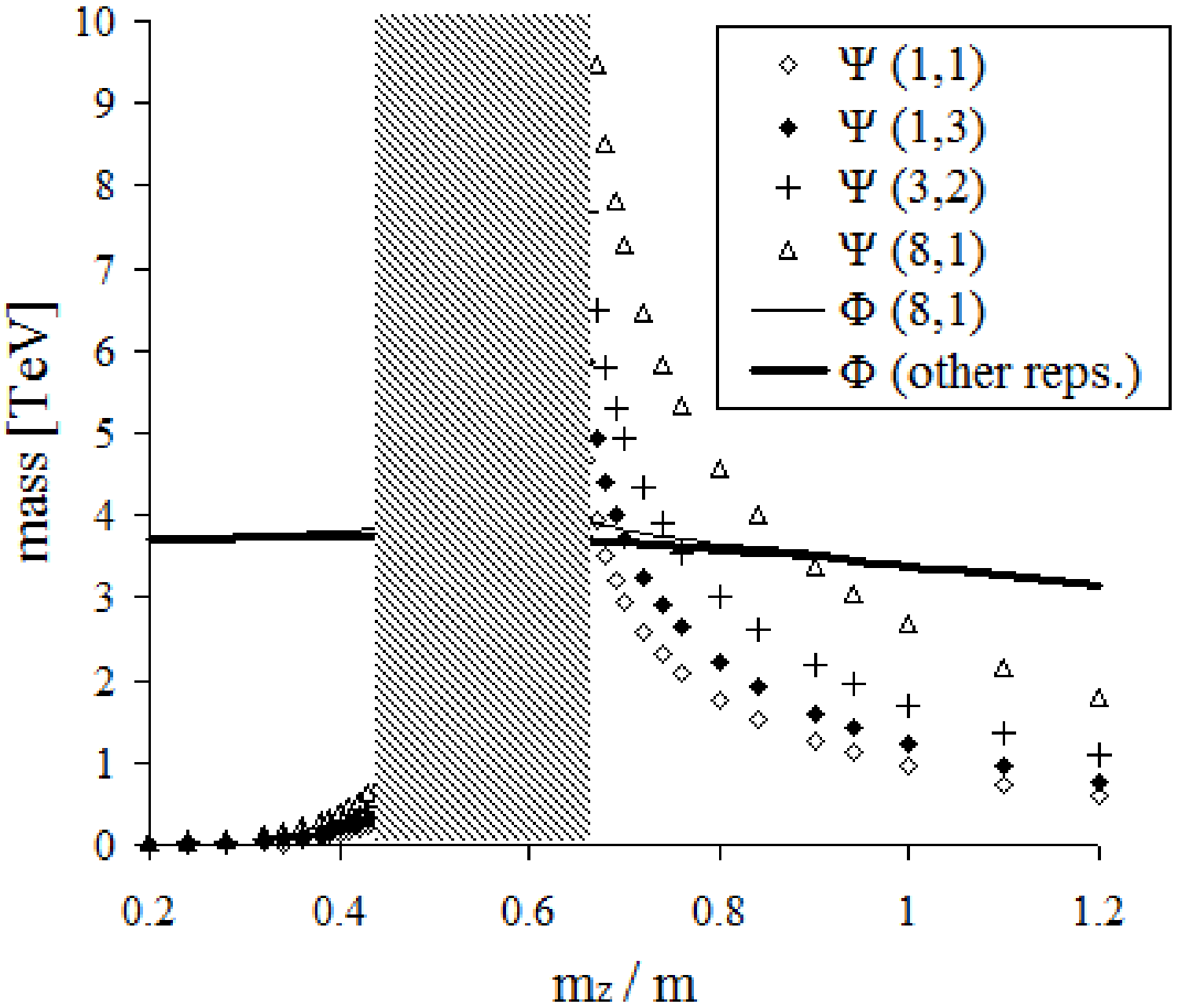}
\caption{The masses of the visible fields from the supersymmetry breaking sector. \label{fig:PhiMasses}}}

\item{\textit{Tachyonic sleptons}}: When $m_z>600$ TeV the sneutrino becomes tachyonic, thus excluding this part of parameter space. For large $\tan\beta$, the stau can become tachyonic at even lower $m_z$ (plot \ref{fig:SleptonMasses} ).

\DOUBLEFIGURE[t]{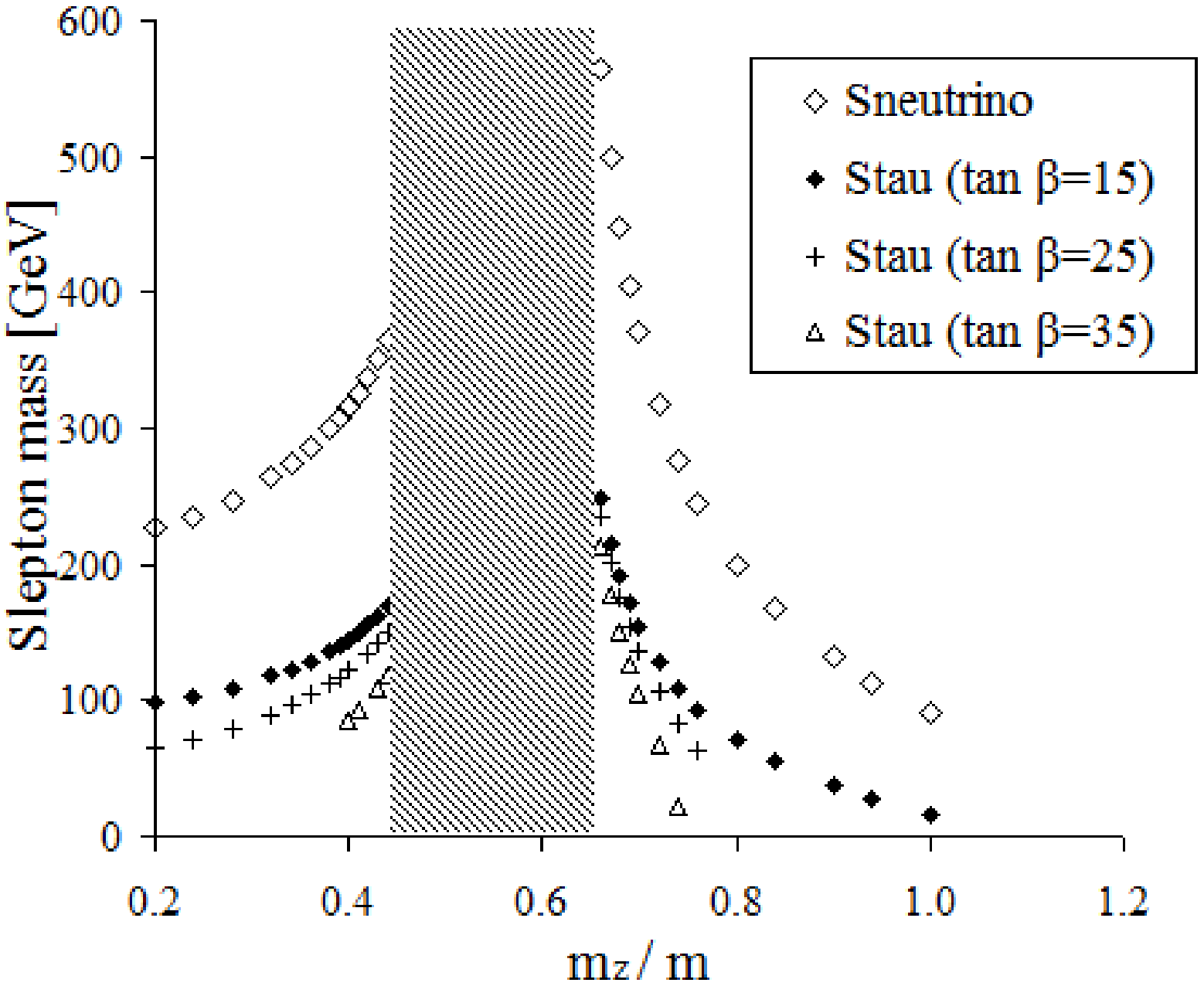,width=8cm}{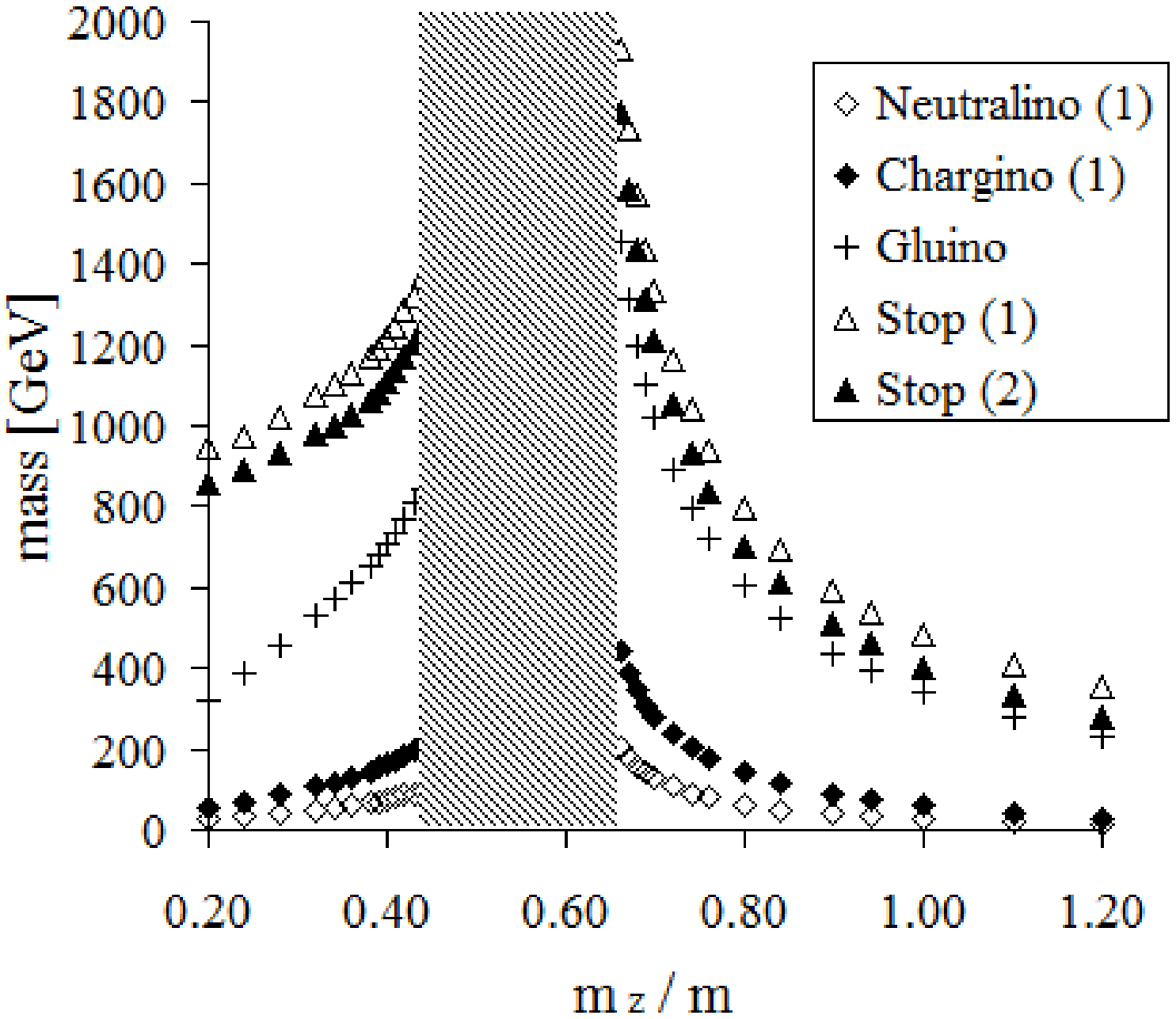,width=8cm}{The slepton masses as a function of $m_z/m$. The stau mass is very sensitive to $\tan\beta$. The sneutrino becomes tachyonic at $m_z/m > 1.2$.\label{fig:SleptonMasses}}{Masses of several sparticles in the spectrum as a function of $m_z/m$. \label{fig:spectra}}

\item{\textit{Light Higgs mass}}: The Higgs mass is in the range $100-117$ GeV, and the LEP bound of
$m_{Higgs}> 114.4$ \cite{Yao:2006px} rules out a large part of parameter space. A large Higgs mass requires large values of $\Lambda_s$, and following the discussion above the allowed region will be where the messengers are lighter, namely $m_z\sim 200$ TeV or $m_z\sim350$ TeV. Also, this constraint excludes $\tan\beta < 5$ (see plots \ref{fig:Higgsmz} and \ref{fig:HiggsMassLambdaS}).

\DOUBLEFIGURE{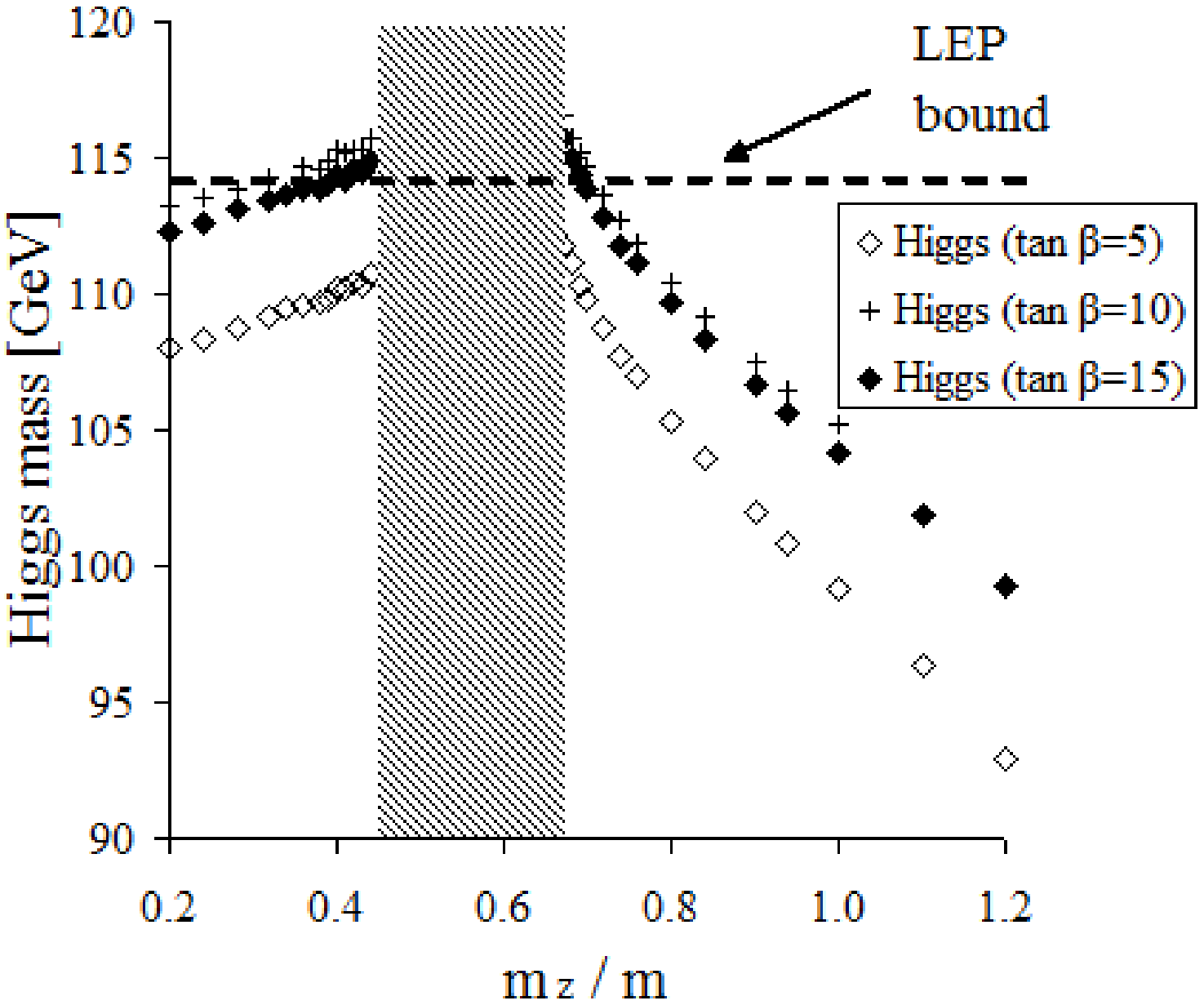,width=8cm}{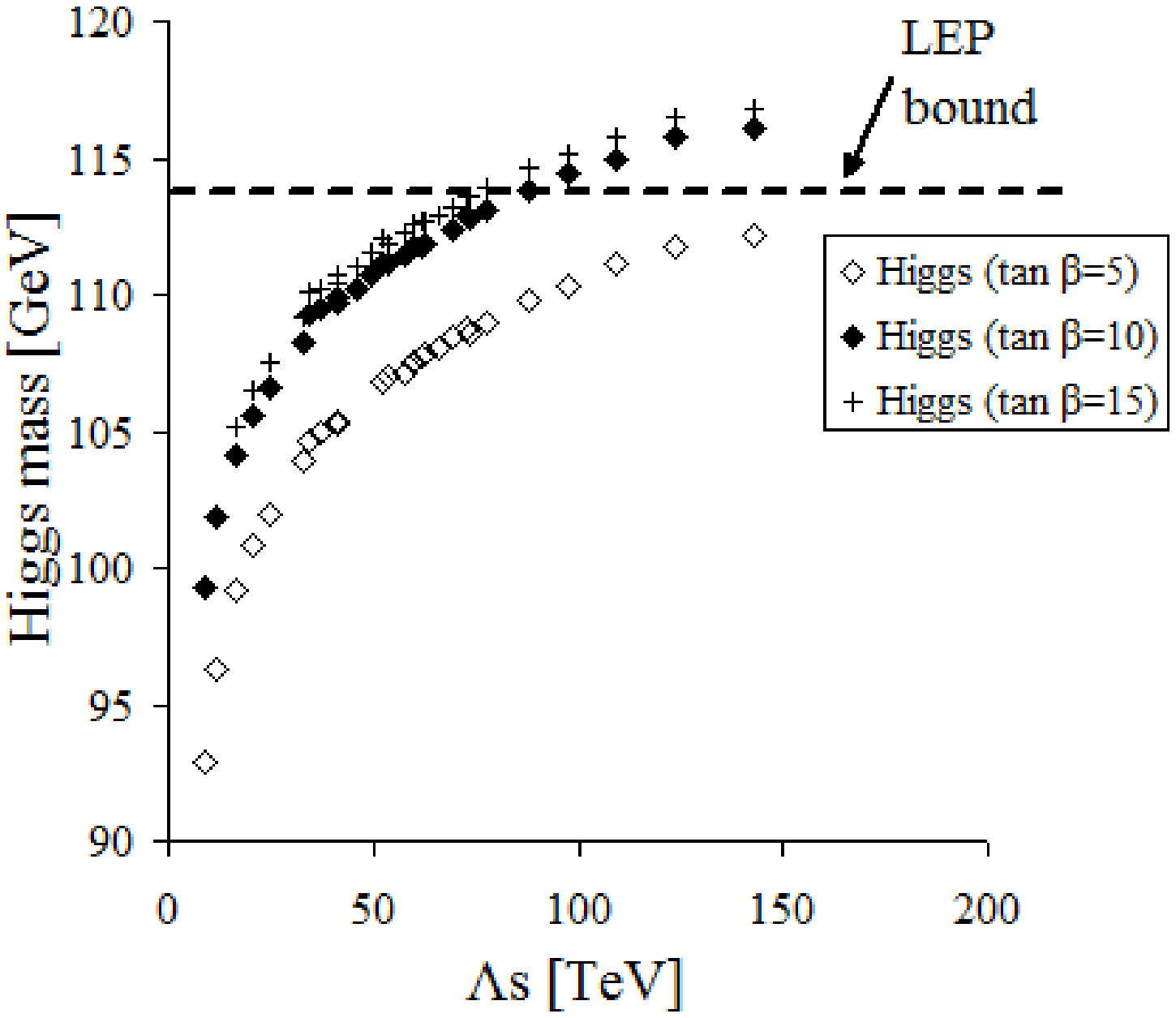,width=8cm}{The Higgs mass as a function of $m_z/m$: The LEP bound rules out a large region of parameter space.\label{fig:Higgsmz}}{The Higgs mass as a function of $\Lambda_s$.\label{fig:HiggsMassLambdaS}}

\item{\textit{Gaugino/scalar mass ratio ($N_{eff}\sim1$)}}: As discussed in section \ref{sec:Gaugino_and_squarks_masses}, the ratio between gaugino masses and scalar masses ($N_{eff}$) is controlled by the parameter $m_z$, and gets values between 0 and 2. However, the range preferred by the Higgs mass constraint, leads to $N_{eff}\sim1$, and no split supersymmetry. Moreover, taking low values of $N_{eff}$ (or equivalently $100\,{\rm TeV} < m_z < 150{\,\rm TeV}$) leads to very light Bino masses, and a neutralino which is lower than 40 GeV.

\item{\textit{$sgn(\mu)$}}: The sign of $\mu$ is a free parameter in GMSB theories, but the different choices lead to similar spectra (In this case the changes are smaller than 1$\%$). The main effect of taking the different signs is a change in the $B_\mu$ parameter, and different chargino and neutralino mixing matrices (the NLSP remains Bino-like). In addition to that, at large $\tan\beta$, where the stau mass is nearly tachyonic, a negative $\mu$ increases the mass, thus increasing slightly the range of allowed parameters.

\item{\textit{The $B\mu/\mu$ problem}}:
The couplings of the Higgs mixing terms, $\mu$ and $B\mu$, are not predicted by the model, but are determined by the values of the $Z$ boson mass and $\tan \beta$.  The computed values of $B\mu/\mu^2$ are approximately proportional to $\tan\beta^{-0.8}$, and are between 0.05 and 0.35 (plot \ref{fig:bmu-mu}). This means that the model has a strong $B/\mu$ problem: in models where the Higgs mixing terms are generated dynamically, this ratio is expected to be at the order of $16\pi^2$ -- namely 2-3 orders of magnitude larger.
%
\end{itemize}

\FIGURE[b]{\includegraphics[width=6.5cm,height=5.5cm]{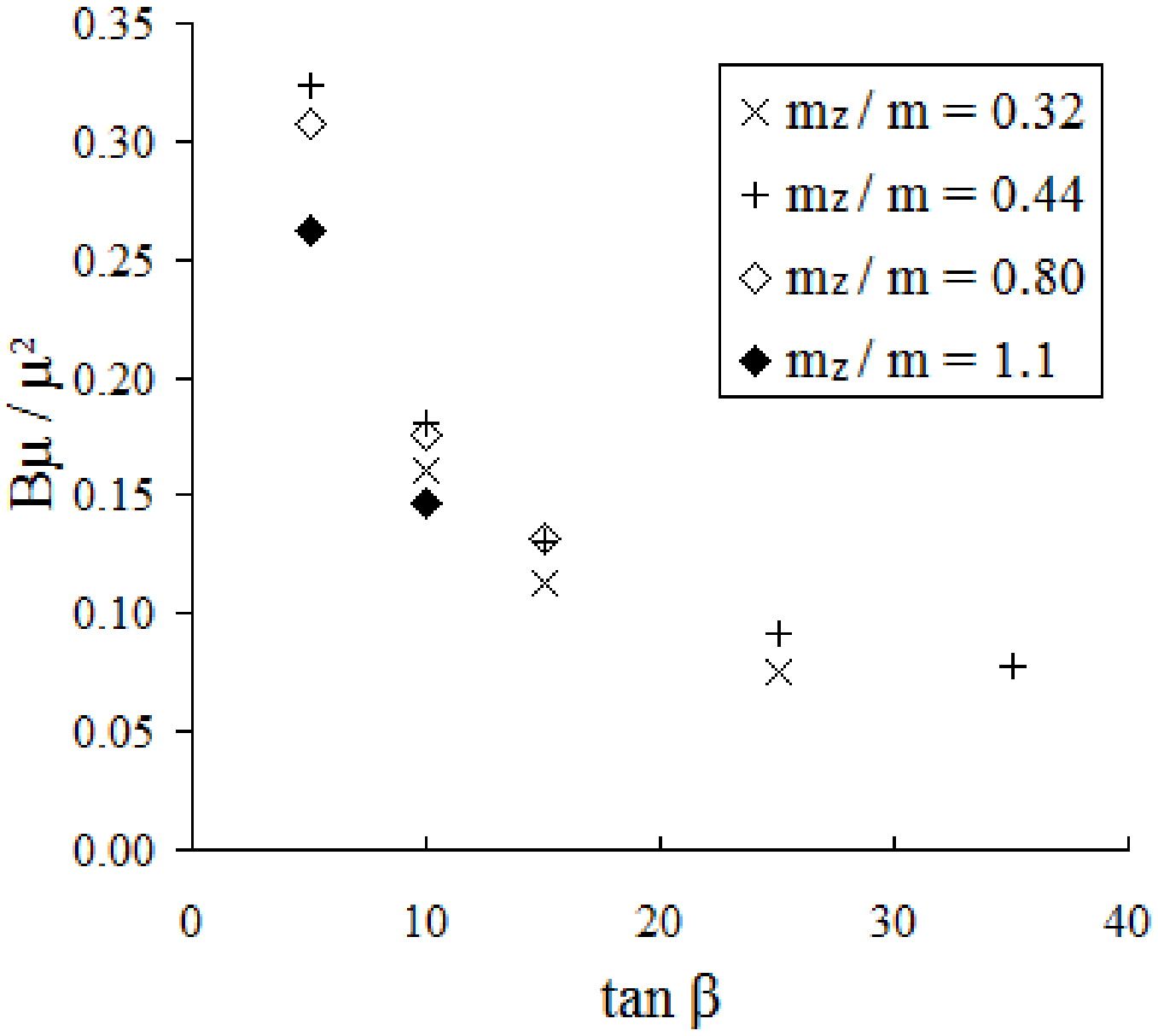}
\caption{$B\mu/\mu^2$ as a function of $\tan\beta$ for several values of $\frac{m_z}{m}$.\label{fig:bmu-mu}}}

Varying the parameters in the allowed ranges discussed in section \ref{constraintson}, for instance by taking a different Yukawa within the range $1<h<2$, leads to similar spectra.
The main difference is the lower masses of the $\Psi_\Phi$ fermions. In the range where the
Higgs mass satisfies the LEP bound, these masses remain above $1$ TeV. By decreasing $h$, the
constraints on the values of $m_z$ change: the excluded window where the light messenger becomes tachyonic moves to larger values of
$m_z$.

\section{A duality cascade in the UV}
\label{cascading}

In models of direct mediation in which the supersymmetry breaking sector is a deformation of ISS there is a tension between a light gravitino with $m_{3\over2}<16$ eV and
gauge coupling unification. To satisfy the first requirement, one needs a supersymmetry breaking scale below a hundred TeV. On the other hand, the supersymmetry breaking sector contains a large number of fields, charged under the MSSM gauge groups, which drive the running couplings towards a Landau pole before reaching unification.
This happens in our model as well.
We will consider a UV completion in terms of a duality cascade \cite{Klebanov:2000hb}.\footnote{A different UV completion has been proposed by \cite{Kitano:2006xg}.} The idea of completing the MSSM
with a duality cascade in the UV is typical of some string theory embedding of the MSSM with D-branes at singularities. Examples of MSSM cascades have been recently presented in \cite{Heckman:2007zp} and \cite{Franco:2008jc}. Typically, one needs to couple an extra sector to the MSSM in order to trigger the first step of the cascade. When we embed the MSSM into a direct mediation model, the cascade is triggered naturally above a certain scale, due to the presence of extra fields charged under the MSSM gauge groups (the messengers), that drive the QCD coupling to a Landau pole.

Let us RG evolve our model to the UV. \DOUBLEFIGURE[t]{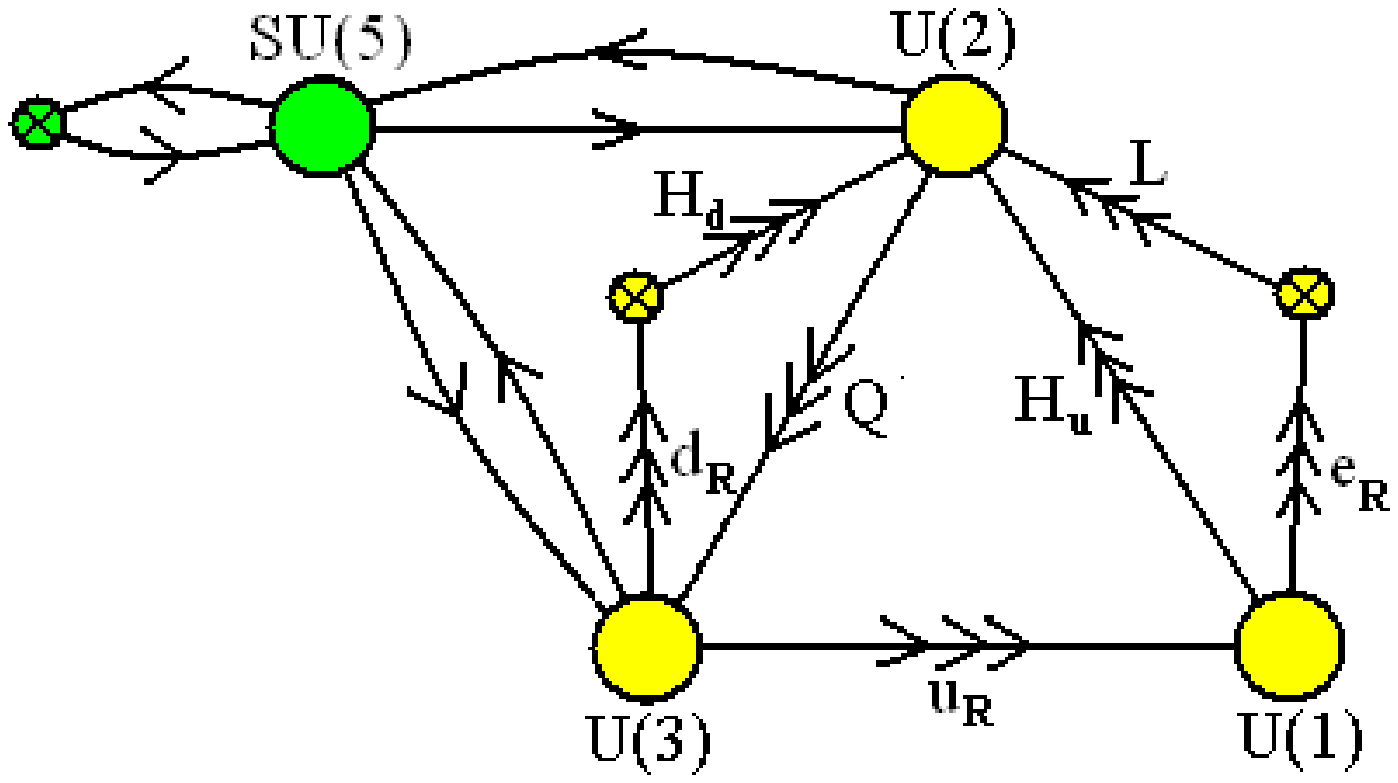,width=7.5cm,height=4.6cm}{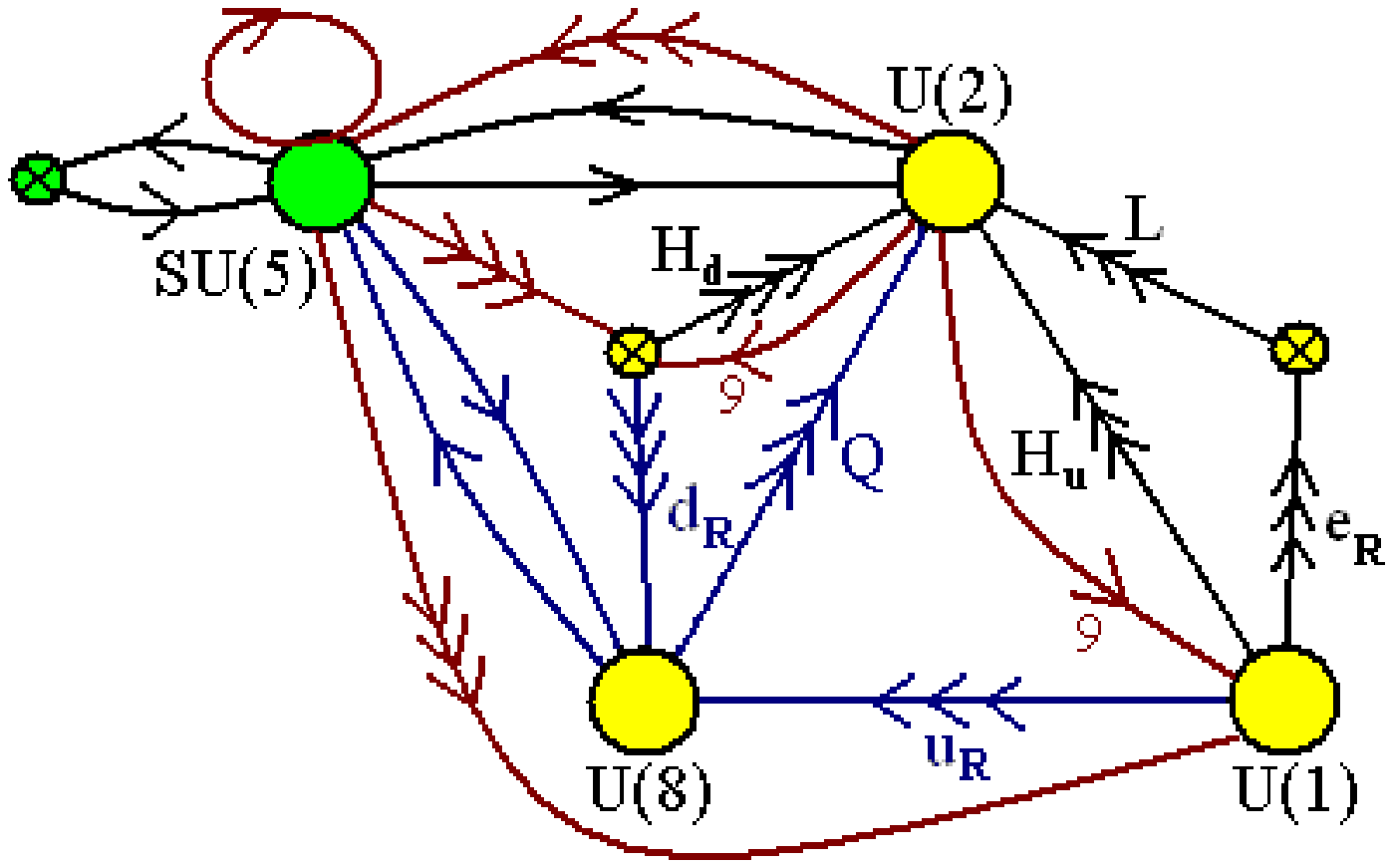,width=8.5cm,height=5cm}{Embedding of the MSSM (yellow) coupled to the supersymmetry breaking sector (green) into a minimal quiver. The adjoints of $SU(2)$ and $SU(3)$ coming from the light DSB sector are not drawn. \label{MSSMquiver}}{The first step of the cascade, after dualizing the QCD node of the MSSM from $SU(3)$ to $SU(8)$ (in red). Dual mesons are red, dual quarks blue, the adjoints of $SU(2)$ and $SU(8)$ are not drawn.\label{MSSMdualquiver}}
Consider first the supersymmetry breaking sector. Above the magnetic cutoff scale $\Lambda_m$, the supersymmetry breaking sector becomes strongly coupled and undergoes a Seiberg duality \cite{Seiberg:1994pq}. Its weakly coupled description is in terms of a $SU(N_c)$ SQCD with $N_f$ light flavors $Q^f$ and $\tilde Q_f$, with $N_c=5$ and $N_f=6$, and a quartic superpotential coupling $W={1\over M}\Tr(\tilde Q_a Q^1)\Tr (Q^a\tilde Q_1)$, which corresponds to the magnetic operator $\Tr\tilde Z Z$ in (\ref{superiss}). The electric description in terms of the quartic coupling is valid up to the scale $M=\Lambda_m^2/ m_z$, which is equal to the GUT scale if we take the magnetic cutoff at $\Lambda_m=5\times10^{7}$ TeV.\footnote{The precise relation is $M=\Lambda_e^3/\Lambda_m m_z$, where $\Lambda_e$ is the dynamical scale of the electric theory. We can take $\Lambda_e\sim \Lambda_m$ up to incalculable coefficients of order one.}

Let us RG evolve the MSSM towards the UV. To properly understand the duality pattern we will consider a minimal embedding of the MSSM and  the supersymmetry breaking sector into a quiver gauge theory as shown in figure \ref{MSSMquiver}.\footnote{We thank Sebastian Franco for discussions on this point.}
We embed the MSSM into a quiver with three nodes \cite{Aldazabal:2000sa}, which has a simple string theory realization with D-branes at a $\mathbb{C}^3/\mathbb{Z}_3$ singularity.\footnote{This quiver is slightly different from the MSSM in two aspects. The first is the presence of two extra anomalous $U(1)$ gauge bosons, however they will get a large mass through the usual Green-Schwarz mechanism for anomaly cancelation. Second, there are two extra pairs of Higgs doublets. A superpotential mass for the Higgs is forbidden by the global $U(1)$ symmetries, however in string theory these symmetries will be explicitly broken by nonperturbative effects. We assume that one can generate an appropriate $\mu$ term for the light Higgses and a large mass for the two extra Higgs pair.}
The green nodes on the left correspond to the electric description of the supersymmetry breaking sector above $\Lambda_m$.
The running of the MSSM couplings is large, due to the extra matter contribution from the supersymmetry breaking sector (see figure \ref{MSSMrunnings}) and the $SU(3)$ coupling hits a Landau pole below the GUT scale at around $10^{9}$ TeV, while all the other couplings are still perturbative.
This triggers a Seiberg duality on the $SU(3)$ node, which has $11$ flavors. In the dual quiver, in figure \ref{MSSMdualquiver}, the dual of the QCD node is an asymptotically free theory with $8$ colors and $11$ flavors (in red). This is the first step of the duality cascade, as we schematically depicted in figure \ref{MSSMdualrunnings}.
The MSSM matter content changes after the duality and, while the dual QCD node is weakly coupled, the weak $SU(2)$ becomes strongly coupled soon triggering the second step of the cascade. The duality cascade will then proceed as discussed in \cite{Heckman:2007zp} and \cite{Franco:2008jc}. The ranks of the gauge groups and the matter content increase fast as one climbs up the UV cascade and at some energy below the GUT scale the field theory description of the system will presumably break down and be replaced by an appropriate string description as in \cite{Klebanov:2000hb}. The issue of unification is still open, it is not unconceivable that the running couplings unify at some point in the UV cascade.

\DOUBLEFIGURE[t]{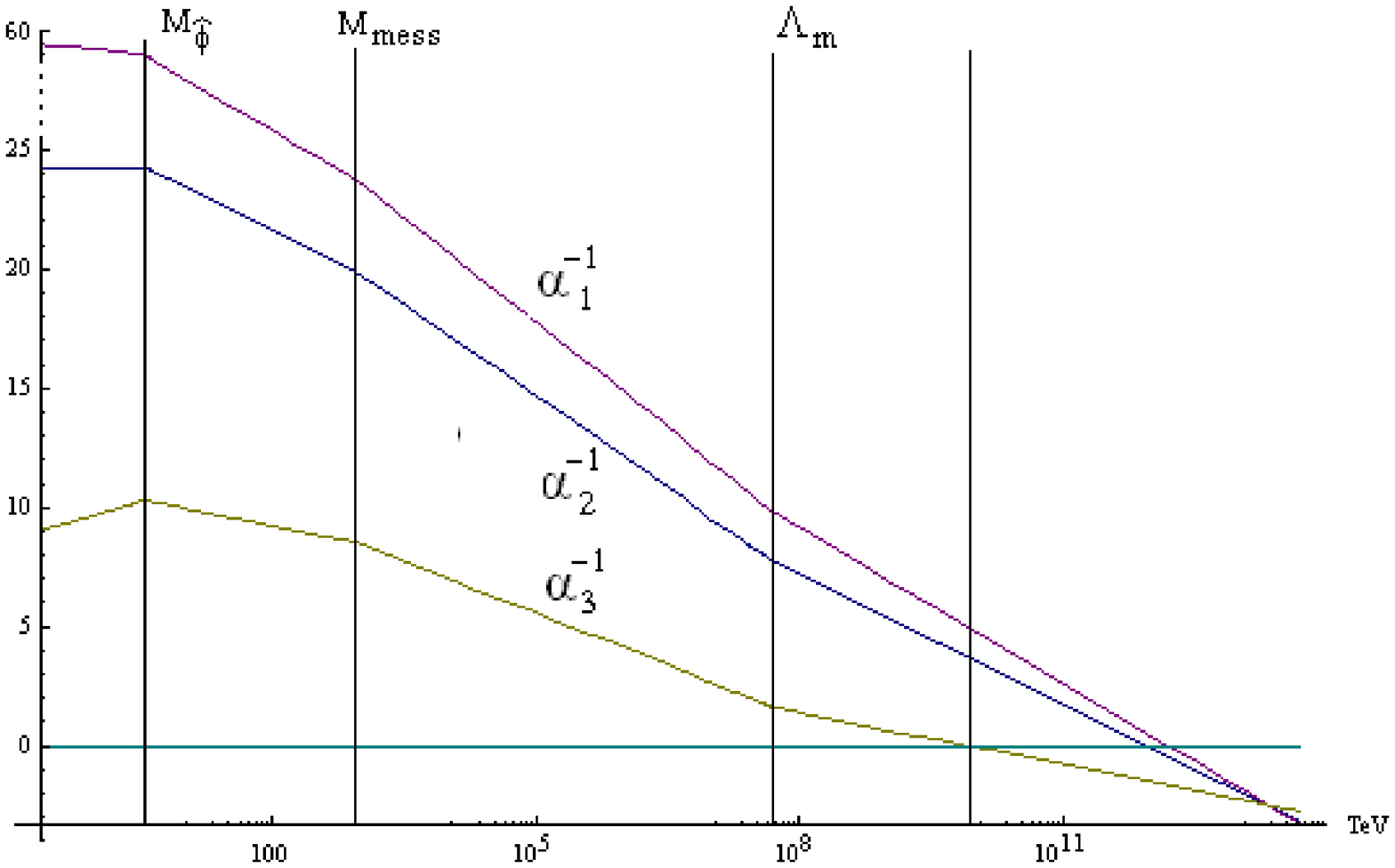,width=7.6cm,height=5.2cm}{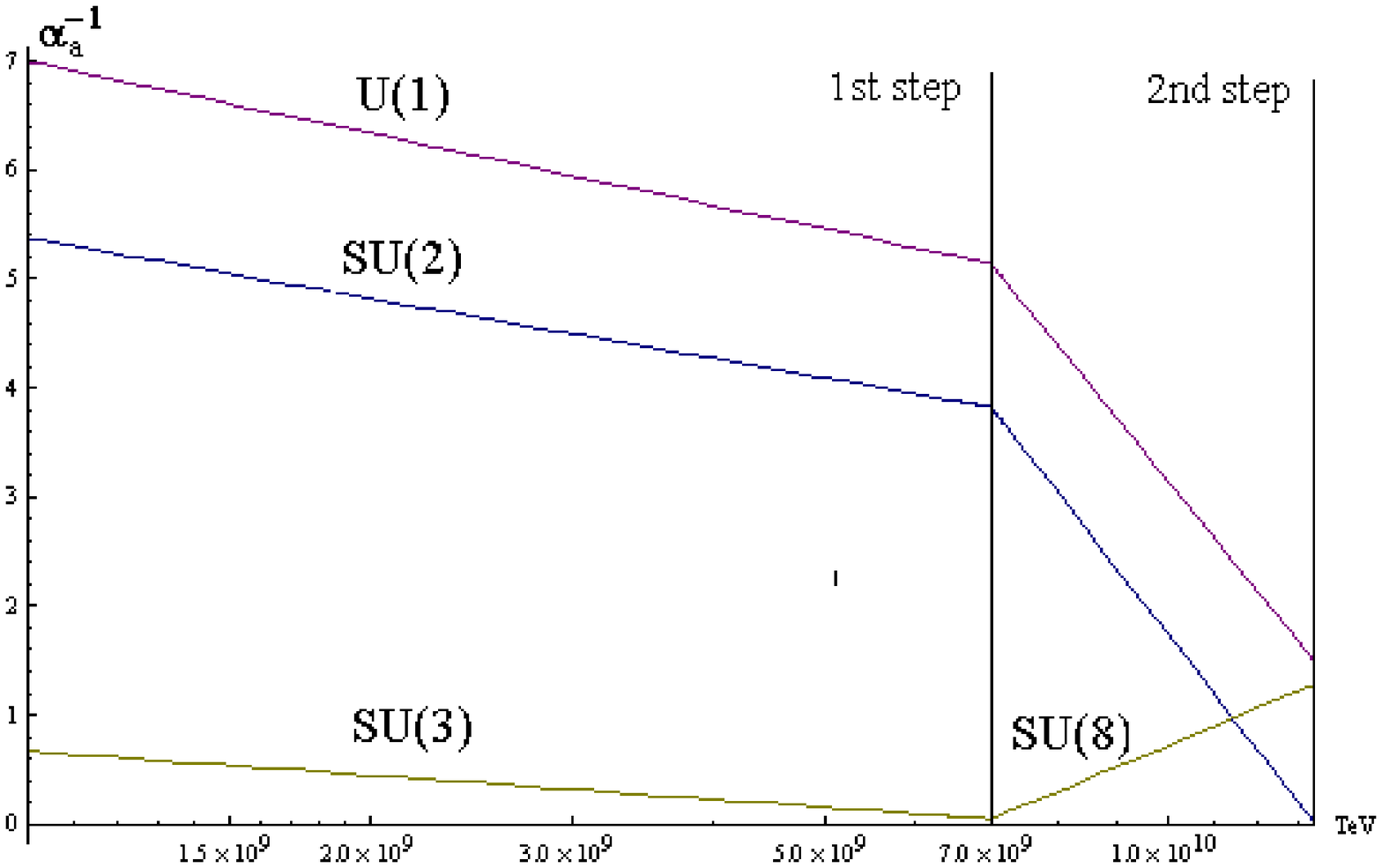,width=7.6cm,height=5.2cm}{ MSSM running couplings with the various thresholds. $\alpha_3$ hits a Landau pole at $0.8\times10^{10}$ TeV, triggering a duality cascade. The MSSM couplings formally unify at the GUT scale at negative values.\label{MSSMrunnings}}{The first step of the cascade, where the QCD node to $SU(8)$ is dualized. In the second step of the cascade the $SU(2)$ is dualized.\label{MSSMdualrunnings}}

\section{Discussion}

We have presented a detailed phenomenology of direct gauge mediation using a deformation of the ISS vacuum, with explicitly broken R-symmetry. One of the aims of this model has been to show that it is indeed possible to obtain a natural MSSM spectrum starting from an ISS vacuum, and we have found on the way new interesting distinctive signatures of this model: an ultralight gravitino, compatible with the cosmological bounds; a light DSB sector, which might be accessible at LHC energies; and long lived DSB sector particles, which might result in cold dark matter candidates. We proposed a UV completion in terms of a duality cascade that will eventually lead to a full string theory description presumably below the GUT scale. The issue of unification is not resolved and deserves further investigation.

One might modify the present model (\ref{super}) by adding all the renormalizable operators allowed by the global symmetries, namely the additional superpotential terms $\delta W_{ren}=g\Tr\hat \Phi^2+f\Tr \hat \Phi^3$. In this case the magnetic theory would be generic, in the sense of \cite{Nelson:1993nf}. Another modification, inspired by string theory constructions, is obtained by adding all the quartic superpotential terms in the electric theory \cite{Giveon:2007ew}, namely $\delta W_{GK}=h^2m_{\hat\Phi}\Tr \hat\Phi^2+h^2m_Y\Tr Y^2$. In both cases, the effect of such operators would be twofold: first, they introduce new classical supersymmetric vacua coming in from infinity. This will reintroduce the tension between long lifetime of the ISS vacuum and non-vanishing gaugino masses that we avoided in our model, pointing towards an unnatural split supersymmetric spectrum. On the other hand, the tree level mass term for the pseudomodulus $\hat\Phi$ will raise the light DSB sector particle masses, probably rendering them inaccessible at LHC energies.

We would like to briefly compare our phenomenology with other models of direct gauge mediation obtained as deformations of ISS, to highlight similarities and differences (see the summary in Table \ref{comparison}. The first four models break explicitly R-symmetry, while the last one breaks it spontaneously.

\noindent {\bf KOO model.} In the original paper of \cite{Kitano:2006xg}, a different embedding of the MSSM gauge group into the ISS flavor symmetry group was considered, namely into $SU(N)_{\rm diag}$. In this case, one needs at least $N_c=11$ and $N_f=16$. To achieve perturbative unification, one needs to push the messenger scale $hm$ as high as $10^{10}\,{\rm TeV}$ and in turn the F-term $F=h\mu^2$ is around $10^{11}\,{\rm TeV}^2$, in order to get soft scalar masses of a few hundred GeV. With such a supersymmetry breaking scale, one gets a gravitino mass $m_{3/2}\sim 50\,{\rm MeV}$, outside the cosmological bounds of \cite{Viel:2005qj}.
The pseudogoldstone boson coming from the fluctuations of ${\rm Re}( \chi-\tilde\chi^T)$ and its fermionic superpartner may give a light supersymmetry breaking sector, with particles in the fundamental representations of the MSSM gauge groups, unlike our case in which the light DSB sector is in the adjoint and bifundamental (they will be beyond the reach of the next colliders though). One may get rid of such light DSB sector particles by gauging the $U(1)_B$ baryon symmetry.

\noindent {\bf Adding singlets.} One can modify the ISS theory by adding extra singlets \cite{Csaki:2006wi}, with explicit R-symmetry breaking superpotential interactions. The pseudomodulus gets an expectation value and the gaugino masses are generated at cubic order in the F-term. Since $F/m_{mess}^2\sim1$, in this case the gaugino masses will be of the same order of the sfermion masses, giving a natural spectrum, however the lightest messenger is very light and can turn tachyonic. The supersymmetry breaking scale is around $100\,{\rm TeV}$, which gives a gravitino mass of order $10\,{\rm eV}$. The theory has a Landau pole for the QCD coupling below the GUT scale, and a UV completion in terms of a duality cascade is suggested as well \cite{Csaki:2006wi}\cite{Csaki:2005fc}.\footnote{In a slight variation of the same model, R-symmetry can be spontaneously broken by the extra singlets.} The low energy spectrum is similar to the one that we discussed in this paper, with light particles coming from the fluctuations of the pseudomodulus $\hat\Phi$, with a mass of a few TeV and the same quantum numbers as in (\ref{decompose}).

\noindent {\bf Adding mesons.} One can explicitly break R-symmetry by adding a quartic superpotential coupling in the magnetic quarks $\delta W=(\tilde q q)^2$ as in \cite{Haba:2007rj}. In this case as well, the gaugino masses are generated only at cubic order in the F-term and the ratio sfermion/gaugino masses is around a hundred, giving a split supersymmetric spectrum. The supersymmetry breaking scale is of order $10^3\,{\rm TeV}$, yielding a gravitino mass around $1\,{\rm keV}$, ruled out by the cosmological bounds \cite{Viel:2005qj}. To get a lighter gravitino, one might lower the supersymmetry breaking scale,  but at some point the messengers become tachyonic and destabilize the vacuum.

\noindent {\bf Adding baryons.} For the particular choice $N_f=7$, $N_c=5$, the magnetic gauge group is $SU(2)$ and one can add a renormalizable operator to the superpotential in the form of a magnetic baryon \cite{Abel:2007jx}\cite{Abel:2007nr}. This theory is generic and achieves a spontaneous radiative breaking of the accidental R-symmetry of the ISS vacuum. In this case one realizes a split supersymmetric spectrum, in which the sfermions are a hundred times heavier than the gauginos. The R-axion is consistent with cosmological bounds, however the supersymmetry breaking scale is around $10^4\,{\rm TeV}$, which gives a gravitino mass of around $50\,{\rm KeV}$, not consistent with the cosmological bounds \cite{Viel:2005qj}. This model has always a Landau pole in the MSSM gauge couplings below the GUT scale.\\

\TABULAR[t]{|c c c c c c c|}{\hline\label{comparison}
{model} & ${R\hspace{-0.25cm}/\hspace{0.25cm}}$&{ MSSM} & $M_{\rm susy} $&$ m_{3/2}$& UV & { DSB sector}  \\\hline
KOO \cite{Kitano:2006xg} & {\rm explicit} & { natur.} & $10^5\,{\rm TeV} $& { $\times$} & GUT & { heavy} \\
{ singlets} \cite{Csaki:2006wi} & {\rm explicit} & { natur.} & $10^2\,{\rm TeV}$ & { $\checkmark$} & cascade& { light} \\
{ mesons} \cite{Haba:2007rj} & {\rm explicit} & { split} & $10^3\,{\rm TeV}$ & { $\times$} & GUT & { light} \\
{ baryons} \cite{Abel:2007jx} & {\rm spont.} & { split} & $10^4\,{\rm TeV}$ & { $\times$} & pole & { light} \\
{ ours} &{\rm explicit} &  { natur.} & $10^2\,{\rm TeV}$ & { $\checkmark$} & cascade & { light}\\ \hline
}{Summary of direct gauge mediation models based on ISS deformations.}

\acknowledgments We would like to thank Nima Arkani-Hamed,  Sebastian Franco,
 Sunny Itzhaki and David Shih for valuable discussions. This work is supported in part by the Israeli Science Foundation center of excellence,
by the Deutsch-Israelische Projektkooperation, by the US-Israel Binational Science
Foundation and
by the European Network.

\appendix

\section{One-loop potential}

The ISS vacuum (\ref{issvev}-\ref{massvev}) has a pseudomodulus $\hat\Phi$. In this appendix, we compute the one loop effective potential for the scalar pseudomodulus $\hat\Phi$, which gives it a mass and an expectation value, as well as the one loop mass for the fermionic superpartner of the pseudomodulus $\psih$.

\subsection{Scalar mass}

Let us compute the masses of the bosons and fermions that couple to $\hat\Phi$ and get splitted by its F-term.
The mass matrix mixes $(\rho,Z,\tilde\rho^\dagger,\tilde Z^\dagger)$ and $(\rho^\dagger, Z^\dagger, \tilde \rho, \tilde Z)$. Its eigenvalues are
 \bea\label{massbo}
 m^2_{1,\pm}=&{h^2\over2}\left(2m^2+h^2m_z^2+\vert\hat\Phi\vert^2
 -\mu^2\pm\sqrt{4m^2(hm_z+\hat\Phi)(hm_z+\hat\Phi^*)+(h^2m_z^2
-\vert\hat\Phi\vert^2+\mu^2)^2}\right) \ ,\cr
 m^2_{2,\pm}=&{h^2\over2}\left(2m^2+h^2m_z^2+\vert\hat\Phi\vert^2
 +\mu^2\pm\sqrt{4m^2(hm_z+\hat\Phi)(hm_z+\hat\Phi^*)+(h^2m_z^2
-\vert\hat\Phi\vert^2-\mu^2)^2}\right) \ ,\cr&
 \eea
while the fermionic eigenvalues are obtained by (\ref{massbo}) setting the F term $\mu^2$ to zero
\be\label{massfe}
{\cal M}_\pm=\left\vert\half h \left(h m_z+\hat\Phi_0 \pm\sqrt{4 m^2+(-h m_z+\hat\Phi_0 )^2}\right)\right\vert \ ,
\ee
There no tachyons if
 \be\label{notac}
 \vert m^2\pm hm_z\hat\Phi\vert^2>\mu^2(m^2+h^2m_z^2) \ .
 \ee
The Coleman-Weinberg potential is
 \bea\label{cw}
 V^{(1)}(\hat\Phi)=&{1\over64\pi^2}\left(\Tr m^4_B\log{m^2_B\over\Lambda_0}-\Tr m^4_F\log{m^2_F\over\Lambda_0}\right)
 \eea
We use the expressions (\ref{massbo}) and (\ref{massfe}) just computed. The lengthy one loop expression can be expanded at first order in the small parameter $\mu/m$
 \bea\label{oneloopo}
 M^2_{\hat\Phi}=&{N\over 8\pi^2} {h\mu^4\over m_Z^3(4m^2+h^2m_Z^2)^{7\over2}}f(h,m,m_Z) \ , \cr
 \hat\Phi_0=&{hm_Z(4m^2+h^2m_Z^2)g(h,m,m_Z)\over 4m^2f(h,m,m_Z)} \ ,
 \eea
where
\bea\label{effe}
f(h,m,m_Z)=&hm_Z\sqrt{4m^2+h^2m_Z^2}
 (h^2m_Z^2-m^2)(m^2+h^2m_Z^2)(2m^2+h^2m_Z^2)+
 \cr&+ m^2(2m^6+12h^2m^4m_Z^2+9h^4m^2m_Z^4+2h^6m_Z^6)\cdot\cr
 &\cdot\log{2m^2+h^2m_Z^2+hm_Z\sqrt{4m^2+h^2m_Z^2}\over 2m^2+h^2m_Z^2-hm_Z\sqrt{4m^2+h^2m_Z^2}} \ , \cr
 g(h,m,m_Z)=&hm_Z\sqrt{4m^2+h^2m_Z^2}
 (-4m^6+10h^2m_Z^2m^4+6h^4m_Z^4m^2+h^6m_Z^6)+
 \cr&+ 2m^4(2m^4-2h^2m^2m_Z^2-h^4m_Z^4)\log{2m^2+h^2m_Z^2+hm_Z\sqrt{4m^2+h^2m_Z^2}\over 2m^2+h^2m_Z^2-hm_Z\sqrt{4m^2+h^2m_Z^2}} \ ,
 \eea
We can further expand at first order in $h$, or at first order in $m_z$:
 \be\label{massvev}
 M^2_{\hat\Phi}={N\over 48\pi^2}{h^4\mu^4\over m^2},\quad
 \hat\Phi_0={h m_z \over2}\ ,
 \ee
where the mass term reproduces the familiar one loop correction to the O'Raifeartaigh model.

\subsection{Fermion mass}

The pseudomodulus $\hat\Phi$ has a fermionic superpartner $\psih$. Its mass is proportional to the vev of the pseudomodulus and it is obtained by integrating out the heavy messengers through the Yukawa interaction
 \be\label{yukawa}
 {\cal L}\supset-h\Tr \psi_\rho\psih\tilde\rho-h\Tr\psih\psi_{\tilde \rho}\rho +{\rm h.c.} \ .
 \ee
This interaction can generate two kind of mass terms at one loop. The Dirac mass $\bar\psih\psih$ vanishes but the Majorana mass term is non-vanishing: roughly speaking it is proportional to the F-term times the expectation value of the pseudomodulus $\hat\Phi$ . In the original ISS, the accidental R symmetry forces $\langle\hat\Phi\rangle=0$ hence the Majorana mass vanishes, while in our case R symmetry is explicitly broken and the mass is non-zero.

\FIGURE[t]{
  \begin{picture}(355,106) (25,-34)
    \SetWidth{0.5}
    \Text(25,11)[lb]{\Large{\Black{$\Psi_{\hat\Phi}$}}}
    \SetWidth{0.5}
    \ArrowArc(105,16)(30,-180,0)
    \DashArrowArc(105,16)(30,-0,180){10}
    \ArrowLine(135,16)(165,16)
    \Text(170,11)[lb]{\Large{\Black{$\Psi_{\hat\Phi}$}}}
    \Text(200,11)[lb]{\Large{\Black{$\Psi_{\hat\Phi}$}}}
    \DashArrowArc(285,16)(30,-0,180){10}
    \ArrowLine(225,16)(255,16)
    \ArrowLine(315,16)(345,16)
    \ArrowLine(45,16)(75,16)
    \ArrowArc(285,16)(30,-180,0)
    \Text(280,-34)[lb]{\Large{\Black{$\Psi_{\tilde\rho}$}}}
    \Text(100,-34)[lb]{\Large{\Black{$\Psi_\rho$}}}
    \Text(280,56)[lb]{\Large{\Black{$\rho$}}}
    \Text(100,56)[lb]{\Large{\Black{$\tilde\rho$}}}
    \Text(350,11)[lb]{\Large{\Black{$\Psi_{\hat\Phi}$}}}
  \end{picture}\caption{
One loop mass $M_\psi$ of the fermionic partner $\psih$ of
the pseudomodulus $\hat\Phi$
\label{loopi}}
}

We need to evaluate the one loop diagrams in figure \ref{loopi}.
We switch from the interaction eigenstates $S=\{\rho,Z,\tilde\rho^\dagger,\tilde Z^\dagger\}$ and $\Psi=\{\psi_\rho,\psi_Z\}$, $\tilde \Psi=\{\psi_{\tilde\rho},\psi_{\tilde Z}\}$ to the mass eigenstates by using the fermionic mass matrix ${\cal M}$ in (\ref{massma}) and the bosonic mass matrix $m^2$ and we introduce the following mixing matrices
 \bea\label{mixmass}
 \widehat m^2=Q^\dagger m^2 Q \ ,\qquad \widehat {\cal M}=U^\dagger {\cal M} V \ ,\cr
 \hat S=SQ \ ,\quad \hat\psi_+=\psi U \ ,\quad \hat\psi_-=\tilde\psi V^* .
 \eea
where $\widehat m^2$ and $\widehat{\cal M}$ are diagonal matrices whose entries are the bosonic and fermionic messenger mass eigenvalues. The interaction term reads
 \be\label{intera}
 {\cal L}\supset -h\Tr\left(\hat\psi_{+p}U^\dagger_{p1}\psih Q_{3i}\hat S^\dagger_i+\psih\hat\psi_{-p}V_{1p}\hat S_iQ^\dagger_{i1}\right) +{\rm h.c.}
 \ee
and the mass term
\be\label{massterm}
{\cal L}\supset -\half M_{\psih}\Tr\psih\psih+{\rm h.c.} \ ,
\ee
is given by the loop integral\footnote{The mass does not actually depend on the cutoff scale $\Lambda$ that we inserted to regulate the integral, due to the unitarity of the mixing matrices.} in figure \ref{loopi}
 \bea\label{diagram}
 M_{\psih}=&-4h^2\sum_{p=1}^4\sum_{i=1,2}Q_{3i}Q^\dagger_{i1}V_{1p}U^\dagger_{p1}
 \int{d^4k\over(2\pi)^4}{\widehat {\cal M}_p\over p^2+\widehat {\cal M}_p^2}{1\over p^2+\widehat m_i^2} \cr
  =& {h^2\over4\pi^2}\sum_{p=1}^4\sum_{i=1,2}Q_{3i}Q^\dagger_{i1}V_{1p}U^\dagger_{p1}
 {\widehat{\cal M}_p\over \widehat m_i^2-\widehat{\cal M}_p^2}
 \left(\widehat m_i^2\ln{\widehat m_i^2\over \Lambda^2}-\widehat{\cal M}_p^2\ln{\widehat{\cal M}_p^2\over \Lambda^2}\right) \ .
 \eea
The mixing matrices can be written using 3 angles
\bea
V_{11}U^\dagger_{11}=\cos^2\theta_f &\qquad&
V_{12}U^\dagger_{21}=\sin^2\theta_f\nonumber \\
Q_{31}Q^\dagger_{11}=-\frac{1}{2}\cos^2\theta_{s1} &\qquad&
Q_{32}Q^\dagger_{21}=\frac{1}{2}\cos^2\theta_{s2}\nonumber \\
Q_{33}Q^\dagger_{31}=-\frac{1}{2}\sin^2\theta_{s1}&\qquad&
Q_{34}Q^\dagger_{41}=\frac{1}{2}\sin^2\theta_{s2}
\eea
where we defined
\bea
\tan 2 \theta_{f}&=&\frac{2m(hm_z+\hat\Phi)}{(hm_z)^2-|\Phi|^2} \nonumber \\
\tan 2 \theta_{s1}&=&\frac{2m(hm_z+\hat\Phi)}{(hm_z)^2-|\Phi|^2+\mu^2} \nonumber \\
\tan 2 \theta_{s2}&=&\frac{2m(hm_z+\hat\Phi)}{(hm_z)^2-|\Phi|^2-\mu^2}
\eea
and the mass eigenstates are ordered from the lightest to the heaviest.

\section{Lifetime}

We evaluated numerically the bounce action for the decay of the ISS vacuum into the closest supersymmetry breaking vacuum (\ref{othervev}) (namely the $n=1$ vacuum). We consider the classical plus one-loop potential $V=\sum_i \vert F_i\vert^2+V_{1-loop}$,
where the F-terms come from (\ref{super}) and the one-loop correction is given in (\ref{oneloop}) and (\ref{oneloopo}). To simplify the computation, we consider a toy model with a slice of the full potential in which we identify $\rho=\tilde \rho$, $Z=\tilde Z$ and we neglect the $\chi,\tilde \chi$ fields, which are fixed to (\ref{issvev}) in both vacua and play no role.
We consider a real slice of the potential so that we are left with a function of four real variables
 \bea\label{potenti}
 V(\rho,Z,Y,\hat\Phi)=&2h^2(Z\rho+mY)^2+2h^2(\hat\Phi\rho+mZ)^2
 +2h^2(hm_ZZ+m\rho)^2\\
 &+h^2(\rho^2-\mu^2)^2+M_{\hat\Phi}^2(\hat\Phi-\hat\Phi_0)^2 \ .\eea

For $m_Z$ inside the phenomenological range (\ref{ConstrainedParameters}), this function has three extrema, one corresponding to the ISS vacuum, the second corresponding to the closest supersymmetry breaking vacuum, and the third giving the saddle point that the bounce crosses in the trajectory between the two vacua. We can plot a one dimensional slice of this potential by the following procedure. Around the ISS vacuum, the lightest fluctuation is $\hat\Phi$, whose mass arises only at one-loop. Hence, we can integrate out the massive fields $\rho,Z,Y$ on their equations of motion coming from (\ref{potenti}) and obtain an effective potential for the real $\hat\Phi$ field only, whose plot is given in figure \ref{potential3D}.

The ISS vacuum is on the plateau on the left, while the other supersymmetry breaking vacuum is on the right. The difference between the value of the potential at the extremum and the ISS value is very small, compared to the difference between the potential at the two vacua, so we can reliably use the triangle approximation to evaluate the bounce action. The peak in the bounce trajectory is reached at
 \be\label{extremum}
 \hat\Phi={m^2-\mu\sqrt{m^2+h^2m_Z^2}\over hm_Z} \ .
 \ee
We evaluate numerically the bounce action and we see that, for $m_Z<260\tev$, the action is larger than the critical value $S_{bounce}\sim400$ for $m_Z<225\tev$.
In the range $330\tev<m_Z<700\tev$, on the other hand, the profile of the potential is reversed, still keeping the same shape: the lower energy vacua (\ref{othervev}) approach the origin of field space, while the ISS vacuum is located at $\hat\Phi_0$, which takes larger values. In this case the bounce action is always very large, $S_{bounce}>>400$. Hence, the requirement that the metastable vacuum be long lived further constrains our parameter space to
\be\label{bouncecoA}
 m_Z<225\tev\quad{\rm or}\quad 330\tev<m_Z<700\tev \ .
\ee

\section{RG flow}
\label{rgfloww}

We list here the RG equation for the masses of the DSB sector light fields  (\ref{visiphi}) coming from the components of the pseudomodulus $\hat\Phi$ and its superpartner $\psih$. In the computation of the low energy spectrum we need to include their running because their mass is one loop suppressed with respect to the messenger mass, hence it is of the order of a TeV. Using the general formulae in the conventions of \cite{Martin:1997ns} we find
 \bea\label{rge}
 (4\pi)^2\beta_{m^2(\phi_i)}=&-8\sum_{a=1,2,3}g_a^2C^a(\phi_i)\vert M_a\vert^2 +{6\over5}g_1^2Y_i{\cal S} \ ,\cr
 (4\pi)^2\beta_{M(\Psi_i)}=&-6 M(\Psi_i)\sum_{a=1,2,3}g_a^2C^a(\psi_i)\ ,
 \eea
where ${\cal S}$ is the trace of the soft masses, weighted by the hypercharge.\footnote{There would normally be a contribution from the Yukawa coupling $h$, but its threshold is  above the messenger mass. This contribution goes like $\log\frac{M_+}{M_-}$, where $M_\pm$ are the messenger masses,
and was neglected.} Note that we have to add to the usual ${\cal S}_{MSSM}$ in eq. (5.57) of \cite{Martin:1997ns} the contribution $\delta {\cal S}$ from the soft masses of the light DSB sector fields that have non-vanishing hypercharge, namely $\tilde p,p$ in (\ref{visiphi}). The total expression appearing in (\ref{rge}) is thus ${\cal S}={\cal S}_{MSSM}+\delta {\cal S}$ where $\delta {\cal S}= 5m_{\tilde p'}^2-5m_{\tilde p}^2$. Note that the fermion masses run faster because their $\beta$ function is proportional to the $\Psi_{\Phi}$ mass, and not the gaugino mass.

The RGE in the MSSM have to be modified accordingly, by replacing ${\cal S}_{MSSM}$ with the full ${\cal S}_{MSSM}+\delta {\cal S}$. The MSSM charges of the various DSB sector fields are collected in the following table. The messengers $(\rho,Z)$ are in the $\bf \bar 5$ of $SU(5)$ and $(\tilde \rho,\tilde Z)$ are in the $\bf 5$. We split them as $\rho=(\rho_3,\rho_2)$ and so on, while the other fields come from (\ref{visiphi})
 \bea
 \matrix{ \rho_2,\,Z_2 & \tilde \rho_2,\,\tilde Z_2 & \rho_3,\,Z_3 & \tilde \rho_3,\,\tilde Z_3 & \varphi_8 & \varphi_3 & \tilde p & \tilde p\,' & S\cr
 ({\bf 1},{\bf 2})_{-1/2} & ({\bf 1},{\bf 2})_{1/2} & ({\bf \bar3},{\bf 1})_{1/3} & ({\bf 3},{\bf 1})_{-1/3} & ({\bf 8},{\bf 1})_0 & ({\bf 1},{\bf 3})_0 & ({\bf 3},{\bf2})_{-5/6} & ({\bf \bar 3},{\bf2})_{5/6} & ({\bf 1},{\bf 1})_0
}\nonumber\eea
The contribution of $\hat{\Phi}$ to the $\beta$ function of the gauge couplings was added at scales above their masses.

Unlike ordinary GMSB models, where the model has a single value for the messengers mass, in this model we have two messenger scales. This fact was partially accounted for by taking the heavier messenger mass as the boundary scale, where the soft supersymmetry breaking mass terms were calculated by integrating out only the heavy messenger. The contribution of the lighter messenger was added as a threshold effect at the light messenger mass scale. Between these scales the light messenger, which is a superfield at the fundamental representation of $SU(5)$, contributes to the beta functions of the gauge coupling
and to the scalar masses
\bea
\Delta\beta_{m_i}=\frac{8}{(4\pi)^4}Str(S(r)\mathcal{M}^2)\sum_{a=1}^3 g_a^4C_a(m_i)~.
\eea
(S(r) is the Dynkin index of the messenger).

As mentioned above, the calculation of the low energy MSSM spectrum was performed using the SoftSUSY software\cite{AllanachSOFTSUSY}. However, the discussed model required several important modification for the RG flow due to the multiple messenger scales and the additional visible fields.

\bibliography{susybreak}

\providecommand{\href}[2]{#2}\begingroup\raggedright\begin{thebibliography}{10}

\bibitem{Intriligator:2006dd}
K.~Intriligator, N.~Seiberg, and D.~Shih, {\it {Dynamical SUSY breaking in
  meta-stable vacua}},  {\em JHEP} {\bf 04} (2006) 021,
  [\href{http://arxiv.org/abs/hep-th/0602239}{{\tt hep-th/0602239}}].

\bibitem{Kitano:2006xg}
R.~Kitano, H.~Ooguri, and Y.~Ookouchi, {\it {Direct mediation of meta-stable
  supersymmetry breaking}},  {\em Phys. Rev.} {\bf D75} (2007) 045022,
  [\href{http://arxiv.org/abs/hep-ph/0612139}{{\tt hep-ph/0612139}}].

\bibitem{Csaki:2006wi}
C.~Csaki, Y.~Shirman, and J.~Terning, {\it {A simple model of low-scale direct
  gauge mediation}},  {\em JHEP} {\bf 05} (2007) 099,
  [\href{http://arxiv.org/abs/hep-ph/0612241}{{\tt hep-ph/0612241}}].

\bibitem{Haba:2007rj}
N.~Haba and N.~Maru, {\it {A Simple Model of Direct Gauge Mediation of
  Metastable Supersymmetry Breaking}},  {\em Phys. Rev.} {\bf D76} (2007)
  115019, [\href{http://arxiv.org/abs/0709.2945}{{\tt arXiv:0709.2945}}].

\bibitem{Abel:2007jx}
S.~Abel, C.~Durnford, J.~Jaeckel, and V.~V. Khoze, {\it {Dynamical breaking of
  $U(1)_{R}$ and supersymmetry in a metastable vacuum}},  {\em Phys. Lett.}
  {\bf B661} (2008) 201--209, [\href{http://arxiv.org/abs/0707.2958}{{\tt
  arXiv:0707.2958}}].

\bibitem{Abel:2007nr}
S.~A. Abel, C.~Durnford, J.~Jaeckel, and V.~V. Khoze, {\it {Patterns of Gauge
  Mediation in Metastable SUSY Breaking}},  {\em JHEP} {\bf 02} (2008) 074,
  [\href{http://arxiv.org/abs/0712.1812}{{\tt arXiv:0712.1812}}].

\bibitem{Dine:2007dz}
M.~Dine and J.~D. Mason, {\it {Dynamical Supersymmetry Breaking and Low Energy
  Gauge Mediation}},  \href{http://arxiv.org/abs/0712.1355}{{\tt
  arXiv:0712.1355}}.

\bibitem{Viel:2005qj}
M.~Viel, J.~Lesgourgues, M.~G. Haehnelt, S.~Matarrese, and A.~Riotto, {\it
  {Constraining warm dark matter candidates including sterile neutrinos and
  light gravitinos with WMAP and the Lyman- alpha forest}},  {\em Phys. Rev.}
  {\bf D71} (2005) 063534, [\href{http://arxiv.org/abs/astro-ph/0501562}{{\tt
  astro-ph/0501562}}].

\bibitem{Klebanov:2000hb}
I.~R. Klebanov and M.~J. Strassler, {\it {Supergravity and a confining gauge
  theory: Duality cascades and chiSB-resolution of naked singularities}},  {\em
  JHEP} {\bf 08} (2000) 052, [\href{http://arxiv.org/abs/hep-th/0007191}{{\tt
  hep-th/0007191}}].

\bibitem{Heckman:2007zp}
J.~J. Heckman, C.~Vafa, H.~Verlinde, and M.~Wijnholt, {\it {Cascading to the
  MSSM}},  {\em JHEP} {\bf 06} (2008) 016,
  [\href{http://arxiv.org/abs/0711.0387}{{\tt arXiv:0711.0387}}].

\bibitem{Franco:2008jc}
S.~Franco, D.~Rodriguez-Gomez, and H.~Verlinde, {\it {N-ification of Forces: A
  Holographic Perspective on D- brane Model Building}},
  \href{http://arxiv.org/abs/0804.1125}{{\tt arXiv:0804.1125}}.

\bibitem{Giveon:2008wp}
A.~Giveon, A.~Katz, and Z.~Komargodski, {\it {On SQCD with massive and massless
  flavors}},  \href{http://arxiv.org/abs/0804.1805}{{\tt arXiv:0804.1805}}.

\bibitem{Nelson:1993nf}
A.~E. Nelson and N.~Seiberg, {\it {R symmetry breaking versus supersymmetry
  breaking}},  {\em Nucl. Phys.} {\bf B416} (1994) 46--62,
  [\href{http://arxiv.org/abs/hep-ph/9309299}{{\tt hep-ph/9309299}}].

\bibitem{Dimopoulos:1996ig}
S.~Dimopoulos and G.~F. Giudice, {\it {Multi-messenger theories of
  gauge-mediated supersymmetry breaking}},  {\em Phys. Lett.} {\bf B393} (1997)
  72--78, [\href{http://arxiv.org/abs/hep-ph/9609344}{{\tt hep-ph/9609344}}].

\bibitem{Cheung:2007es}
C.~Cheung, A.~L. Fitzpatrick, and D.~Shih, {\it {(Extra)Ordinary Gauge
  Mediation}},  \href{http://arxiv.org/abs/0710.3585}{{\tt arXiv:0710.3585}}.

\bibitem{Beenakker:1996ch}
W.~Beenakker, R.~Hopker, M.~Spira, and P.~M. Zerwas, {\it {Squark and gluino
  production at hadron colliders}},  {\em Nucl. Phys.} {\bf B492} (1997)
  51--103, [\href{http://arxiv.org/abs/hep-ph/9610490}{{\tt hep-ph/9610490}}].

\bibitem{Tung:2006tb}
W.~K. Tung {\em et~al.}, {\it {Heavy quark mass effects in deep inelastic
  scattering and global QCD analysis}},  {\em JHEP} {\bf 02} (2007) 053,
  [\href{http://arxiv.org/abs/hep-ph/0611254}{{\tt hep-ph/0611254}}].

\bibitem{AllanachSOFTSUSY}
B.~Allanach, {\it Softsusy: A c++ program for calculating supersymmetric
  spectra},  {\em Comput. Phys. Commun.} {\bf 143} (2002) 305,
  [\href{http://arxiv.org/abs/hep-ph/0104145}{{\tt hep-ph/0104145}}].

\bibitem{Yao:2006px}
{\bf Particle Data Group} Collaboration, W.~M. Yao {\em et~al.}, {\it {Review
  of particle physics}},  {\em J. Phys.} {\bf G33} (2006) 1--1232.

\bibitem{Seiberg:1994pq}
N.~Seiberg, {\it {Electric - magnetic duality in supersymmetric nonAbelian
  gauge theories}},  {\em Nucl. Phys.} {\bf B435} (1995) 129--146,
  [\href{http://arxiv.org/abs/hep-th/9411149}{{\tt hep-th/9411149}}].

\bibitem{Aldazabal:2000sa}
G.~Aldazabal, L.~E. Ibanez, F.~Quevedo, and A.~M. Uranga, {\it {D-branes at
  singularities: A bottom-up approach to the string embedding of the standard
  model}},  {\em JHEP} {\bf 08} (2000) 002,
  [\href{http://arxiv.org/abs/hep-th/0005067}{{\tt hep-th/0005067}}].

\bibitem{Giveon:2007ew}
A.~Giveon and D.~Kutasov, {\it {Stable and Metastable Vacua in Brane
  Constructions of SQCD}},  {\em JHEP} {\bf 02} (2008) 038,
  [\href{http://arxiv.org/abs/0710.1833}{{\tt arXiv:0710.1833}}].

\bibitem{Csaki:2005fc}
C.~Csaki, G.~Marandella, Y.~Shirman, and A.~Strumia, {\it {The super-little
  Higgs}},  {\em Phys. Rev.} {\bf D73} (2006) 035006,
  [\href{http://arxiv.org/abs/hep-ph/0510294}{{\tt hep-ph/0510294}}].

\bibitem{Martin:1997ns}
S.~P. Martin, {\it {A supersymmetry primer}},
  \href{http://arxiv.org/abs/hep-ph/9709356}{{\tt hep-ph/9709356}}.

\end{thebibliography}\endgroup
\bibliographystyle{JHEP}

\end{document}